\documentclass[sigconf,screen]{acmart}

\AtBeginDocument{%
  \providecommand\BibTeX{{%
    \normalfont B\kern-0.5em{\scshape i\kern-0.25em b}\kern-0.8em\TeX}}}

\copyrightyear{2022}
\acmYear{2022}
\setcopyright{acmlicensed}
\acmConference[ESEC/FSE '22]{Proceedings of the 30th ACM Joint European Software Engineering Conference and Symposium on the Foundations of Software Engineering}{November 14--18, 2022}{Singapore, Singapore}
\acmBooktitle{Proceedings of the 30th ACM Joint European Software Engineering Conference and Symposium on the Foundations of Software Engineering (ESEC/FSE '22), November 14--18, 2022, Singapore, Singapore} 
\acmPrice{15.00}
\acmDOI{10.1145/3540250.3549125} 
\acmISBN{978-1-4503-9413-0/22/11}

\usepackage{tcolorbox}
\usepackage{xspace}
\usepackage{stfloats}
\usepackage{enumitem}
\usepackage{booktabs}
\usepackage{makecell}
\usepackage{subcaption}
\usepackage{graphicx}
\usepackage{multirow}
\newcolumntype{C}[1]{>{\centering\arraybackslash}m{#1}}
\usepackage{balance}

\makeatletter
\newcommand\footnoteref[1]{\protected@xdef\@thefnmark{\ref{#1}}\@footnotemark}

\theoremstyle{acmdefinition}
\newtheorem{exmp}{Example}[section]

\newcommand{\congyingEdit}[1]{\textcolor{black}{#1}}

\newcommand{\tocheck}[1]{\textcolor{black}{#1}}
\newcommand{\tool}{\textsc{Tracer}\xspace}

\makeatother

\usepackage{microtype}
\usepackage{fancyhdr}
\fancypagestyle{firstpageheader}{
  \fancyhf{} 
  \fancyhead[C]{Accepted to the 30th {ACM} Joint European Software Engineering Conference and Symposium on the Foundations of Software Engineering}
  \setlength{\headheight}{13.6pt} 
}

\begin{document}
\title{Tracking Patches for Open Source Software Vulnerabilities}

\author{Congying Xu}
\authornote{Also with Shanghai Key Laboratory of Data Science, and Shanghai Collaborative Innovation Center of Intelligent Visual Computing.}
\affiliation{
\department{School of Computer Science}
\institution{Fudan University}
\city{Shanghai}
\country{China}
}

\author{Bihuan Chen}
\authornotemark[1]
\authornote{Bihuan Chen is the corresponding author.}
\affiliation{
\department{School of Computer Science}
\institution{Fudan University}
\city{Shanghai}
\country{China}
}

\author{Chenhao Lu}
\authornotemark[1]
\affiliation{
\department{School of Computer Science}
\institution{Fudan University}
\city{Shanghai}
\country{China}
}

\author{Kaifeng Huang}
\authornotemark[1]
\affiliation{
\department{School of Computer Science}
\institution{Fudan University}
\city{Shanghai}
\country{China}
}

\author{Xin Peng}
\authornotemark[1]
\affiliation{
\department{School of Computer Science}
\institution{Fudan University}
\city{Shanghai}
\country{China}
}

\author{Yang Liu}
\affiliation{
\department{School of Computer Science and Engineering}
\institution{Nanyang Technological University}
\country{Singapore}
}


\begin{abstract}
Open source software (OSS) vulnerabilities threaten the security~of software systems that use OSS. Vulnerability databases provide~valuable~information~(e.g., vulnerable version and patch) to mitigate~OSS vulnerabilities. There~arises~a~growing concern about~the~information quality of vulnerability databases. However, it is unclear~what~the quality of patches~in~existing vulnerability databases is; and existing manual or heuristic-based approaches for patch tracking are either too expensive or too specific to apply to all OSS vulnerabilities.

To address these problems, we first conduct~an empirical~study~to understand the quality and characteristics~of~patches for OSS vulnerabilities in two industrial vulnerability databases. Inspired~by~our study,~we then propose the first automated approach, \tool,~to~track patches for OSS vulnerabilities from multiple knowledge~sources.~Our evaluation has demonstrated that~i)~\tool~can~track patches for~up to \tocheck{273.8\%} more vulnerabilities than heuristic-based~approaches~while achieving a higher F1-score by~up~to~\tocheck{116.8\%};~and~ii) \tool~can~complement industrial vulnerability databases. Our evaluation has also indicated the generality and practical usefulness of \tool.



\end{abstract}



\maketitle

\thispagestyle{firstpageheader}


\section{Introduction}

Open source software (OSS) provides the foundation for open~source and proprietary applications. It allows developers~to~reuse functionalities instead of reinventing the wheel. As revealed~by~a~recent report~\cite{Synopsys-report},~98\% of applications contain OSS.~However, security risks are also introduced with OSS. \congyingEdit{84\% of} applications contain at least an OSS vulnerability in 2020,~a~\congyingEdit{9\%} increase from~the~75\%~in~2019; and each application contains an average of 158 OSS vulnerabilities \cite{Synopsys-report}. Even~worse,~OSS vulnerabilities~are~detected~at an increasing speed, nearly doubling in the last two years~\cite{Snyk-report}.~Hence, OSS vulnerability~management has become more and more urgent.


Great efforts have been made to mitigate~security risks~in~OSS~vulnerabilities. 
Vulnerability databases play~ a~significant role in these~efforts by providing valuable information~(e.g., \congyingEdit{description,} vulnerable versions,~and patches) for different vulnerability analysis tasks.~CVE List \cite{cve} and~NVD~\cite{nvd} are well-known public vulnerability~databases, which even~go~beyond OSS vulnerabilities. They provide data feed~of the entire database, and often serve as the main vulnerability~source of industrial vulnerability databases, e.g., Black Duck~\cite{blackduck},~WhiteSource \cite{whitesource}, Veracode~\cite{veracode} and Snyk~\cite{snyk}. These industrial vulnerability databases are specifically focused on OSS vulnerabilities. 

\textbf{Problem.} While vulnerability databases are accumulating~a~massive collection of vulnerabilities, there arises an increasing concern about information quality of vulnerability databases. Nguyen~and Massacci \cite{nguyen2013reliability} and Dong~et~al. \cite{dong2019towards} revealed~the~unreliability~of~vulnerable version data in vulnerability databases. 
Chaparro et al.~\cite{chaparro2017detecting} and Mu~et al.~\cite{mu2018understanding} showed the prevalence of missing reproducing steps in vulnerability descriptions. The missing or inaccurate information of vulnerability entries in vulnerability databases makes~it challenging to timely mitigate OSS vulnerabilities in applications.

Patch is a valuable piece of information to capture a vulnerability. It enables not only OSS vulnerability mitigation (e.g., software~composition analysis~\cite{ponta2020detection, pashchenko2020vuln4real, Wang2020empirical}), but also other security tasks~(e.g., hot patch generation and deployment~\cite{mulliner2013patchdroid, duan2019automating, xu2020automatic},~patch~presence~testing \cite{zhang2018precise, jiang2020pdiff, dai2020bscout}~and~vulnerability detection~\cite{li2016vulpecker, li2018vuldeepecker, jang2012redebug, kim2017vuddy, xiao2020mvp, cui2020vuldetector}).~On one hand, \congyingEdit{the accuracy of these tasks is affected if vulnerability patches are missing or inaccurate.} \tocheck{However, the problem~is~that~\textit{it is unclear~what~the~quality of patches in vulnerability databases is.}} 

On the other hand,~apart from leveraging patches in vulnerability databases, these tasks track vulnerability patches~by~manual efforts \cite{xu2020automatic, jiang2020pdiff, dai2020bscout, zhou2017automated, sabetta2018practical, chen2020machine, xiao2020mvp, ponta2020detection, pashchenko2020vuln4real},~by heuristic rules~like~looking for commits in CVE references~\cite{duan2019automating, li2016vulpecker, li2018vuldeepecker}~and searching~for~CVE~identifiers in commits~\cite{you2017semfuzz, Wang2020empirical}, or from~security advisories~that~list~the~vulnerabilities and their patches for specific projects~\cite{mulliner2013patchdroid, jang2012redebug, kim2017vuddy}. \tocheck{However, the problem is that \textit{these patch tracking approaches are either~too expensive or too specific to apply to all OSS vulnerabilities.}}


\textbf{Empirical Study.} To address the first problem, we conduct~an~empirical study to understand the quality and characteristics~of~patches for OSS vulnerabilities in two industrial vulnerability databases~\congyingEdit{(i.e., Veracode~\cite{veracode} and Snyk~\cite{snyk})}. On the basis of a dataset~of~\tocheck{10,070}~vulnerabilities, we find that patches are~missing for more than half~of~the vulnerabilities in the two databases;~and patches are inconsistent~across the two databases for around 20\%~of the vulnerabilities. Further,~based on a dataset of \tocheck{1,295} vulnerabilities, we manually locate their accurate patches, and observe that patches are in the form~of~GitHub~commits for around 93\% of the vulnerabilities; and multiple patches~are developed for about 41\% of the vulnerabilities, for which the two databases only have a patch recall of around 50\%.

\textbf{Our Approach.} Inspired by our empirical study, we~propose~an automated approach, \tool, to address the second problem. \tool is designed to automatically track patches~(in the form of commits) for an OSS vulnerability from multiple knowledge sources~(i.e.,~NVD \cite{nvd}, Debian~\cite{debian}, Red Hat~\cite{redhat} and GitHub).
Our key idea~is~that~patch commits~are often frequently referenced during the reporting, discussion and resolution of an OSS vulnerability. \tool~works~in~three steps. First, given the CVE identifier of an OSS vulnerability, it constructs~a reference network based on multiple knowledge sources,~to model resource references~during~vulnerability reporting, discussion and resolution. Second, it selects~patch~commits~in the network that have high connectivity and~high~confidence. Finally,~it~expands the selected~patch~commits~by searching relevant commits across branches of a repository so as to track patches more completely.


\textbf{Evaluation.} To demonstrate the effectiveness of \tool, we~compare~it with three heuristic-based patch tracking approaches~and~two \congyingEdit{industrial} vulnerability databases on the \tocheck{1,295} vulnerabilities used~in our empirical study.
Our~evaluation has indicated that i) \tool can find patches for \tocheck{58.6\%}~to~\tocheck{273.8\%} more vulnerabilities than heuristic-based approaches; ii) for the vulnerabilities whose patches~are~found, \tool can have a higher patch accuracy~by~up to \tocheck{116.8\%}~in~F1-score than heuristic-based approaches;~and iii)~\tool can complement industrial databases by finding patches~completely.

To demonstrate the generality of \tool, we run~it~against \tocheck{3,185} vulnerabilities for which only one of the two industrial vulnerability databases provides their patches and \tocheck{5,468} vulnerabilities for which none of the two industrial vulnerability databases provides their patches. Our evaluation~has~shown that \tool can find patches~for \tocheck{67.7\%} and \tocheck{51.5\%} of the vulnerabilities, and achieve a sampled patch precision of \tocheck{0.823} and \tocheck{0.888} and a sampled patch recall of \tocheck{0.845}~and \tocheck{0.899}. Moreover, to evaluate the practical usefulness~of~\tool, we conduct a user study with 10 participants. Our evaluation has~shown that \tool can help track patches more accurately and quickly.

\textbf{Contribution.} This work makes the following contributions.
\begin{itemize}[leftmargin=*]
\item We conducted a large-scale empirical study to understand~the~quality and~characteristics of patches for OSS vulnerabilities.


\item We proposed the first automated approach, named \tool,~to~track patches of OSS vulnerabilities from multiple knowledge sources.

\item We conducted extensive experiments to demonstrate the~effectiveness, generality and practical usefulness of \tool.
\end{itemize}


\section{An Empirical Study}\label{sec:study}

We design an empirical study to understand the quality and characteristics of patches for \congyingEdit{OSS} vulnerabilities in vulnerability databases by answering the following research questions.

\begin{itemize}[leftmargin=*]
\item \textbf{RQ1 Coverage Analysis:} how many \congyingEdit{OSS} vulnerabilities have patches included in vulnerability databases? (Sec.~\ref{sec:coverage})
\item \textbf{RQ2 Consistency Analysis:} how many \congyingEdit{OSS} vulnerabilities~have consistent patches across vulnerability databases? (Sec.~\ref{sec:consistency})
\item \textbf{RQ3 Type Analysis:} what are the common patch types for \congyingEdit{OSS} vulnerabilities in vulnerability databases?  (Sec.~\ref{sec:type})
\item \textbf{RQ4 Cardinality Analysis:} what are the mapping cardinalities between \congyingEdit{OSS} vulnerabilities and their patches? (Sec.~\ref{sec:cardinality})
\item \textbf{RQ5 Accuracy Analysis:} how is the patch accuracy~of~\congyingEdit{OSS}~vulnerabilities in vulnerability databases? (Sec.~\ref{sec:accuracy})
\end{itemize}

We design \textbf{RQ1} to measure the prevalence of missing patches~for \congyingEdit{OSS} vulnerabilities in different vulnerability databases. We use~\textbf{RQ2} to quantify the prevalence of inconsistent patches~for \congyingEdit{OSS} vulnerabilities across different vulnerability databases. We leverage~\textbf{RQ3}~and \textbf{RQ4} to capture the common patch types~and mapping cardinalities between \congyingEdit{OSS} vulnerabilities and their patches. We develop~\textbf{RQ5}~to~assess the accuracy of patches~for \congyingEdit{OSS} vulnerabilities in different~vulnerability databases. In summary, our results from \textbf{RQ1},~\textbf{RQ2}~and \textbf{RQ5} aim to assess patch quality~from~different perspectives~and~motivate the need~for~an~automated approach to accurately find patches for \congyingEdit{OSS} vulnerabilities, and our results from \textbf{RQ3} and \textbf{RQ4} aim to capture~the~characteristics of patches from different perspectives, and inspire~the~design of our automated patch tracking approach.

\subsection{Data Preparation}\label{sec:preparation}

\textbf{Vulnerability Database Selection.} Official vulnerability databases (e.g., CVE List and NVD) do not provide a ``patch'' field for each~collected vulnerability entry. Industrial vulnerability databases~leverage official vulnerability databases as the main vulnerability~source, and claim that they collect patches manually or semi-automatically. Therefore, industrial vulnerability databases often provide a ``patch'' field for their collected vulnerabilities. To enable large-scale empirical study of patches, we focus on industrial vulnerability databases.



Initially, we selected the vulnerability databases from four~companies, Black Duck~\cite{blackduck}, WhiteSource~\cite{whitesource}, Veracode~\cite{veracode}~and~Snyk~\cite{snyk}. They~provide~software composition analysis to identify~OSS~used~in an application and report any OSS vulnerabilities. Hence,~we~believe they achieve good coverage of \congyingEdit{OSS} vulnerabilities. However,~Black Duck does~not make the vulnerability database publicly accessible, and WhiteSource does not disclose \congyingEdit{patches}~for~each~vulnerability. We finally selected \congyingEdit{Veracode's and Snyk's~vulnerability databases. Hereafter they are referred to as $DB_A$ and $DB_B$.} We also confirmed with Veracode and Snyk that their public databases are complete.

\textbf{Breadth Dataset Construction.} To broadly quantify missing patches in the two industrial vulnerability databases and inconsistent patches~across~them (i.e., \textbf{RQ1} and \textbf{RQ2}), we built~a~breadth dataset of \congyingEdit{OSS} vulnerabilities by crawling all the OSS vulnerabilities from $DB_A$ and $DB_B$ as of April 7, 2020 by their publicly available interfaces. We obtained \tocheck{8,630} and \tocheck{5,858} CVEs from $DB_A$~and~$DB_B$. $DB_A$~and~$DB_B$ contain a union of \tocheck{10,070} CVEs.

\textbf{Depth Dataset Construction.} To accurately characterize patch types, mapping cardinalities and patch accuracy (i.e., \textbf{RQ3}, \textbf{RQ4}~and \textbf{RQ5}), we built a depth dataset of \congyingEdit{OSS} vulnerabilities\congyingEdit{,} whose size is smaller than the breadth dataset but whose patches~are~all manually identified to ensure the completeness and accuracy. Specifically, to balance the ease of patch accuracy \congyingEdit{analysis} across~two~databases and the effort of manual patch identification, we selected~\tocheck{1,417}~CVEs for which \congyingEdit{both databases} reported their patches. For each CVE, two of the authors separately found its patches by analyzing patches~reported by both databases, looking into~CVE description~and~references in NVD, and searching GitHub repositories and Internet~resources. 
\congyingEdit{When they had disagreements, a third author was involved into the discussion for consensuses.} Finally, they successfully found patches for \tocheck{1,295} CVEs, while they were~still~uncertain for \tocheck{122}~CVEs due to limited disclosed~information. 
These \tocheck{1,295} CVEs mainly cover seven programming languages. Therefore, we believe our depth dataset is representative of \congyingEdit{OSS} vulnerabilities.

\subsection{Coverage Analysis (RQ1)}\label{sec:coverage}



\begin{figure}[!t]
\centering
\begin{subfigure}[b]{0.2\textwidth}
\centering
\includegraphics[scale=0.56]{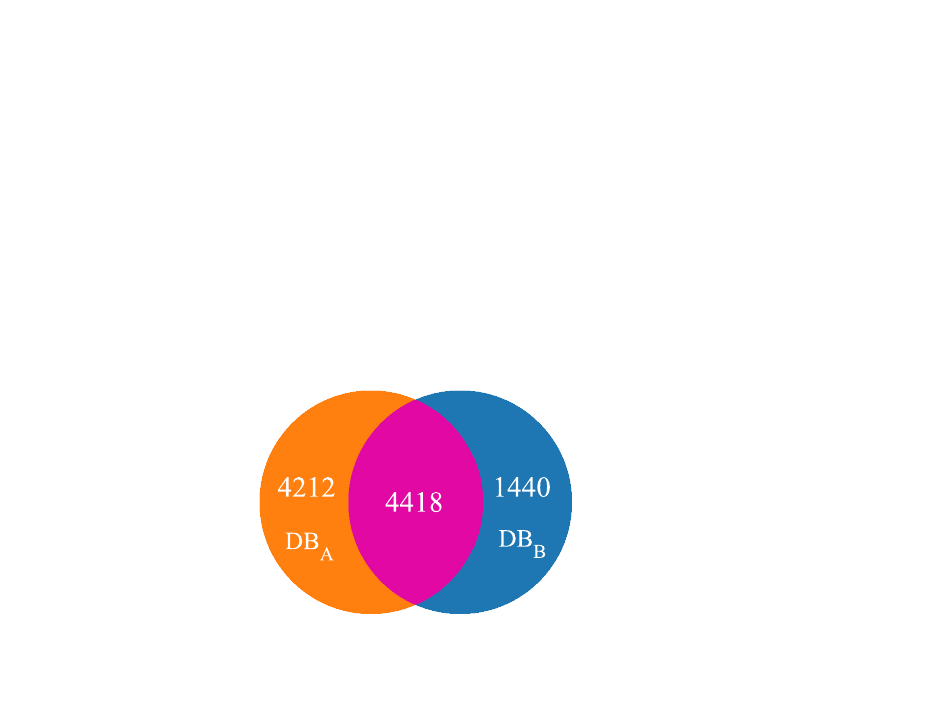}
\vspace{-5pt}
\caption{CVEs}\label{fig:rq1-cves}
\end{subfigure}
~
\begin{subfigure}[b]{0.2\textwidth}
\centering
\includegraphics[scale=0.56]{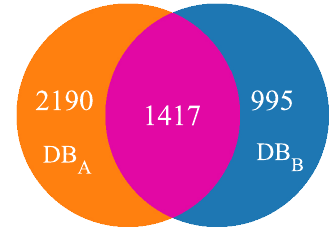}
\vspace{-5pt}
\caption{CVEs with Patches}\label{fig:rq1-cves-with-patches}
\end{subfigure}
\vspace{-10pt}
\caption{Overlap Between Two Databases}\label{fig:intersection}
\end{figure}

Fig.~\ref{fig:rq1-cves} shows the overlap of CVEs between $DB_A$ and $DB_B$,~and~Fig.~\ref{fig:rq1-cves-with-patches} presents the overlap of CVEs with patches between~them. We~can~see that different vulnerability databases have different coverage~of~\congyingEdit{OSS} vulnerabilities. $DB_A$ and $DB_B$ have an overlap of \tocheck{4,418} CVEs, while respectively covering \tocheck{4,212} and \tocheck{1,440} unique CVEs. Moreover, \tocheck{3,607 (41.8\%)} and \tocheck{2,412 (41.2\%)}~of~the CVEs~have their patches provided~in $DB_A$ and $DB_B$. Overall, of all the \tocheck{10,070} CVEs, \tocheck{4,602 (45.7\%)} CVEs have patches provided by at least one vulnerability database. These results~indicate that both vulnerability databases have a moderately low~patch coverage, and missing patches are prevalent. Automated patch tracking approaches are needed to help find missing~patches.

\subsection{Consistency Analysis (RQ2)}\label{sec:consistency}

To analyze patch consistency across the two databases, we focus~on CVEs with patches (i.e., Fig.~\ref{fig:rq1-cves-with-patches}). As a CVE may have~a~set~of~patches, we consider two databases as having consistent patches for a CVE~if their patch~sets are the same.
\congyingEdit{We distinguish patch inconsistency~between existence and content inconsistency.}
The former refers to~two cases~that~one~database provides patches~for~a~CVE, but the other~database either~does~not cover the CVE, or covers~the CVE but does~not provide patches. It reflects the incompleteness of collected \congyingEdit{OSS} vulnerabilities and their patches. The latter refers to two cases~that~both databases provide~patches~for a CVE, and their patch sets have~an inclusion~relationship,~or~do~not have an inclusion relationship~but~are different from each other. It shows the potential inaccuracy~of~patches.

Table~\ref{table:consistency} shows our patch consistency analysis results. The first~column gives the number of CVEs with consistent patches.~The~second to fourth columns report the number of CVEs with existence~inconsistent patches, and last three columns list the number~of~CVEs~with content inconsistent patches. It can be seen that i) only~\tocheck{907~(19.7\%)} of the \tocheck{4,602} CVEs have~consistent patches; ii) more than two-thirds (i.e., \tocheck{3,185 (69.2\%)}) of the CVEs have existence inconsistency, where \tocheck{1,392 (30.2\%)}~of~the~CVEs are not included in $DB_{A}$ or $DB_{B}$, and~\tocheck{1,793 (39.0\%)} of the CVEs~are included but do not have patches in $DB_{A}$~or $DB_{B}$; and iii) \tocheck{510 (11.1\%)} of the CVEs incur content inconsistency, where \tocheck{176 (3.8\%)} of the CVEs' patches~from~one database are included in the patches from the other;~and~\tocheck{334~(7.3\%)}~of~the~CVEs~have different~and~non-inclusive patch sets across $DB_{A}$ and $DB_{B}$. 
These results indicate that these vulnerability databases often report inconsistent patches, the incompleteness of collected \congyingEdit{OSS} vulnerabilities and their patches is severe in these vulnerability databases.

\begin{table}[!t]
\centering
\small
\caption{Patch Consistency and Inconsistency Results}\label{table:consistency}
\vspace{-10pt}
\begin{tabular}{|*{1}{C{2.4em}}|*{1}{C{2.3em}}*{1}{C{3.3em}}*{1}{C{3.8em}}|*{1}{C{2.3em}}*{1}{C{3.3em}}*{1}{C{3.8em}}|}
\noalign{\hrule height 1pt}
\multirow{2}{*}{Cons.} & \multicolumn{3}{c|}{Existence Inconsistency} & \multicolumn{3}{c|}{Content Inconsistency} \\\cline{2-4}\cline{5-7}
 & Total & No CVE & No Patch & Total & Inclusion & Difference \\\noalign{\hrule height 1pt}
907 (19.7\%) & 3,185 (69.2\%) & 1,392 (30.2\%) & 1,793 (39.0\%) & 510 (11.1\%) & 176 (3.8\%) & 334 (7.3\%) \\
\noalign{\hrule height 1pt}
\end{tabular}
\end{table}

\subsection{Type Analysis (RQ3)}\label{sec:type}

We find \tocheck{3,043} patches for the \tocheck{1,295} CVEs in our depth dataset~by~manual analysis. Specifically, \tocheck{2,852 (93.7\%)} patches are in the type~of GitHub commits potentially due to a wide adoption of GitHub~across open source software. \tocheck{136 (4.5\%)} patches are in the type~of~SVN commits potentially due to the prevalence of SVN before the introduction of GitHub, whereas only \tocheck{55 (1.8\%)} patches are in the~type~of commits from other Git platforms. Besides, from the perspective of CVEs, \tocheck{1,202 (92.8\%)} of the \tocheck{1,295} CVEs have the patches~in~the~type~of GitHub commits, \tocheck{4 (0.3\%)} CVEs have their patches in the type~of~SVN commits, and \tocheck{48 (3.7\%)} CVEs have their patches in the type of both GitHub and SVN commits due to the migration from SVN to GitHub. Only \tocheck{30 (2.3\%)} CVEs have some patches in the type of commits~from other Git platforms. These results demonstrate that patches for \congyingEdit{OSS} vulnerabilities are mostly in the type of GitHub~commits.~Thus,~patch tracking approaches can specifically focus on GitHub~commits.

\subsection{Cardinality Analysis (RQ4)}\label{sec:cardinality}


\congyingEdit{
We categorize three~types~of mapping cardinalities between CVEs and their~patches.} 
In~detail, \tocheck{567 (43.8\%)}~of~the CVEs have a \textit{one-to-one mapping} to their patch;~i.e., they have~only one single~patch~to~fix~the vulnerability. Hereafter this category~is~referred to as \textit{SP}.

\tocheck{195 (15.1\%)} of the~CVEs have a \textit{one-to-some mapping} to the~patches, meaning that they~have multiple \textit{equivalent} patch~sets and~any~one~of the patch sets is sufficient to patch the vulnerability. Hereafter~we refer to this category as \textit{MEP}. \congyingEdit{Two patches are equivalent~if~they~have the same code differences.} ~It~is~mainly~caused by two reasons. First,~a CVE is patched by a pull request which~is~merged. Thus,~the~pull~request commits and merged commits~are equivalent patch sets for the CVE. 
Second,~the~repository of OSS~is migrated from~SVN~to~GitHub. Thus, commits for patching a CVE can be in the repository~on~SVN and GitHub, and SVN commits and GitHub commits are equivalent. 

\tocheck{533 (41.2\%)} of the CVEs have a \textit{one-to-many mapping} to patches, which can be further classified into three~types. 
\congyingEdit{First,~a~CVE~is~fixed by multiple separate commits in a branch. This is because the~CVE~is difficult to fix or the initial patch is not complete.}
Hereafter we refer to this type as \textit{MP}, accounting for \tocheck{101 (7.8\%)} of the CVEs. 
Second, a CVE~is~fixed by multiple patch sets in multiple~branches~because the CVE affects~multiple versions of OSS~and~each~version~is~maintained on a separate branch. These multiple patch sets should~be~identified because patches for different versions can be different. Hereafter this type is referred~to~ as \textit{MB}, covering \tocheck{372 (28.7\%)}~of~the~CVEs.~Third, a 
\congyingEdit{CVE is fixed by multiple patch sets in multiple repositories.~This is because the CVE affects multiple OSS, or multiple versions~of~OSS~are maintained in separate repositories.}
Hereafter we refer to this type as \textit{MR}, which covers \tocheck{60 (4.6\%)} of the CVEs. 

\congyingEdit{These results demonstrate various mapping cardinalities~between CVEs~and~their~patches. They should be considered to ensure completeness when tracking patches for OSS vulnerabilities.}

\subsection{Accuracy Analysis (RQ5)}\label{sec:accuracy}

\begin{table}[!t]
\centering
\small
\caption{Patch Accuracy of Two Databases}\label{table:accuracy}
\vspace{-10pt}
\begin{tabular}{|c|c|*{3}{C{2.0em}}|*{3}{C{2.0em}}|}
\noalign{\hrule height 1pt}
\multirow{2}{*}{Cardinality} & \multirow{2}{*}{Number} &  \multicolumn{3}{c|}{$DB_A$} & \multicolumn{3}{c|}{$DB_B$} \\\cline{3-8}
& & Pre. & Rec. & F1 & Pre. & Rec. & F1 \\
\noalign{\hrule height 1pt}
1:1 (SP) & 567       & 0.908 & 0.915 & 0.910   & 0.900 & 0.921 & 0.906   \\
1:$i$ (MEP) & 195    & 0.935 & 0.898 & 0.902  & 0.924 & 0.909  & 0.906   \\
1:$n$ (MP) & 101     & 0.923 & 0.483 & 0.616  & 0.911 & 0.520 & 0.638    \\
1:$n$ (MB) & 372     & 0.941 & 0.510 & 0.620  & 0.932 & 0.436 & 0.555    \\
1:$n$ (MR) & 60      & 0.913 & 0.610 & 0.695  & 0.964 & 0.526 & 0.636   \\\hline
Total & 1,295       & 0.923 & 0.748 & 0.793  & 0.917 & 0.730 & 0.771     \\
\noalign{\hrule height 1pt}
\end{tabular}
\end{table}

We use precision, recall and F1-score as the indicators~of~patch~accuracy. For the CVEs having one-to-some mapping to their patches,~we consider reporting one of the multiple equivalent patches~as~correct. For example, for a CVE that has two equivalent patches,~a~database that reports one of the two equivalent patches has a full~precision and a full recall, while a database that reports one~of~the~two~equivalent patches and another irrelevant patch achieves a half precision and a full recall. Table~\ref{table:accuracy} breaks down the accuracy results~of~the two databases with respect to the mapping cardinalities.
\congyingEdit{The second~column reports the number of CVEs in each mapping cardinality,~and the next six columns~report patch accuracy in the two databases.}

\begin{figure*}[!t]
\centering
\includegraphics[scale=0.47]{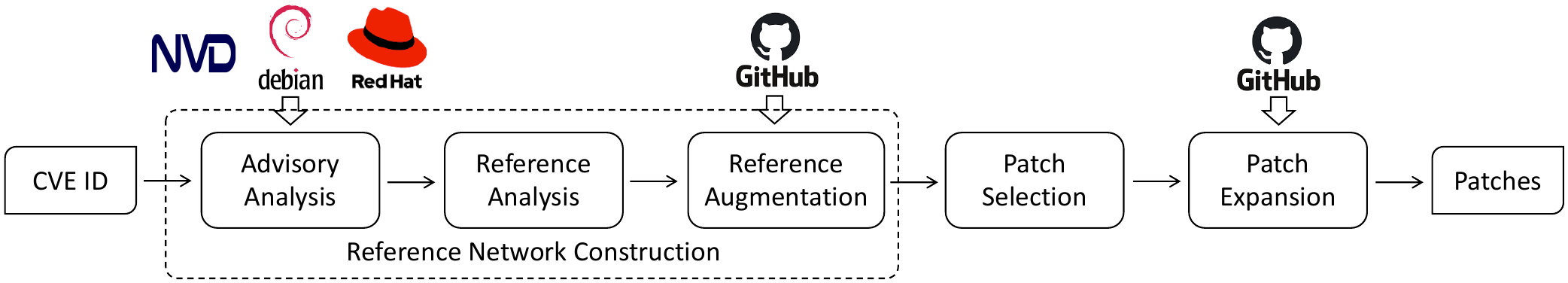}
\vspace{-5pt}
\caption{Approach Overview of \tool}\label{fig:overview}
\end{figure*}

$DB_A$ and $DB_B$ achieve a high~precision and a high recall~of~about 90\% for the CVEs belonging~to~\textit{SP}~and~\textit{MEP}, while having~a~high~precision of above 90\% but a low recall~of~around 50\% for the CVEs~having one-to-many mappings to their patches (i.e., \textit{MP}, \textit{MB} and \textit{MR}). 
These results show that these vulnerability databases often~miss~some patches, especially for CVEs with multiple patches, and such~missing information would make it challenging to achieve accurate~software composition analysis. It reflects the need to automatically find complete patches for \congyingEdit{OSS} vulnerabilities.



\section{Our Approach}


Based on the findings from our empirical study, we propose~an~automated~approach, \tool, to track patches (in the form of commits)~for \congyingEdit{OSS} vulnerabilities.~The underlying idea of \tool is that patch~commits are often frequently referenced during the reporting, discussion and resolution~of~an~\congyingEdit{OSS} vulnerability in various advisory~sources. Fig.~\ref{fig:overview}~presents an overview of \tool. It takes as an input the CVE identifier of an \congyingEdit{OSS} vulnerability, and returns its patches. \tool works in three steps.~First,~it~constructs a reference network for the CVE starting from multiple~advisory sources  (i.e., NVD, Debian~\cite{debian}, Red Hat~\cite{redhat} and GitHub). The~goal~is to model resource references~during~the reporting, discussion and~resolution of the CVE. 
Second, it selects the patch nodes~(i.e., patch commits) in the network~which have high connectivity and high~confidence\congyingEdit{,} and thus are most likely to be patches for the CVE. Finally, it expands the selected patch commits via searching relevant commits across branches of the same repository. The goal is to establish a potential one-to-many~mapping between the CVE and its patches.~In~the~following subsections, we will elaborate \congyingEdit{on} each step in detail.

\subsection{Reference Network Construction}\label{sec:networkConstruction}

The first step of \tool consists of three sub-steps. The first~two~sub-steps (i.e., \textit{advisory analysis} and \textit{reference analysis}) construct~a~reference network via analyzing advisories from NVD,~Debian and Red Hat. The last sub-step (i.e., \textit{reference augmentation}) augments~the~reference network by searching relevant commit links from GitHub.

\textbf{Advisory Analysis.} \tool first initializes the reference~network by setting the CVE under analysis as the \textit{root node}.~It~then~adds three~\textit{advisory source nodes} (i.e., NVD, Debian and Red Hat) as the child~node~of the root node. These advisory source nodes are used~to visualize where the finally selected patches originate.

\begin{figure}[!t]
\centering
\includegraphics[scale=0.4]{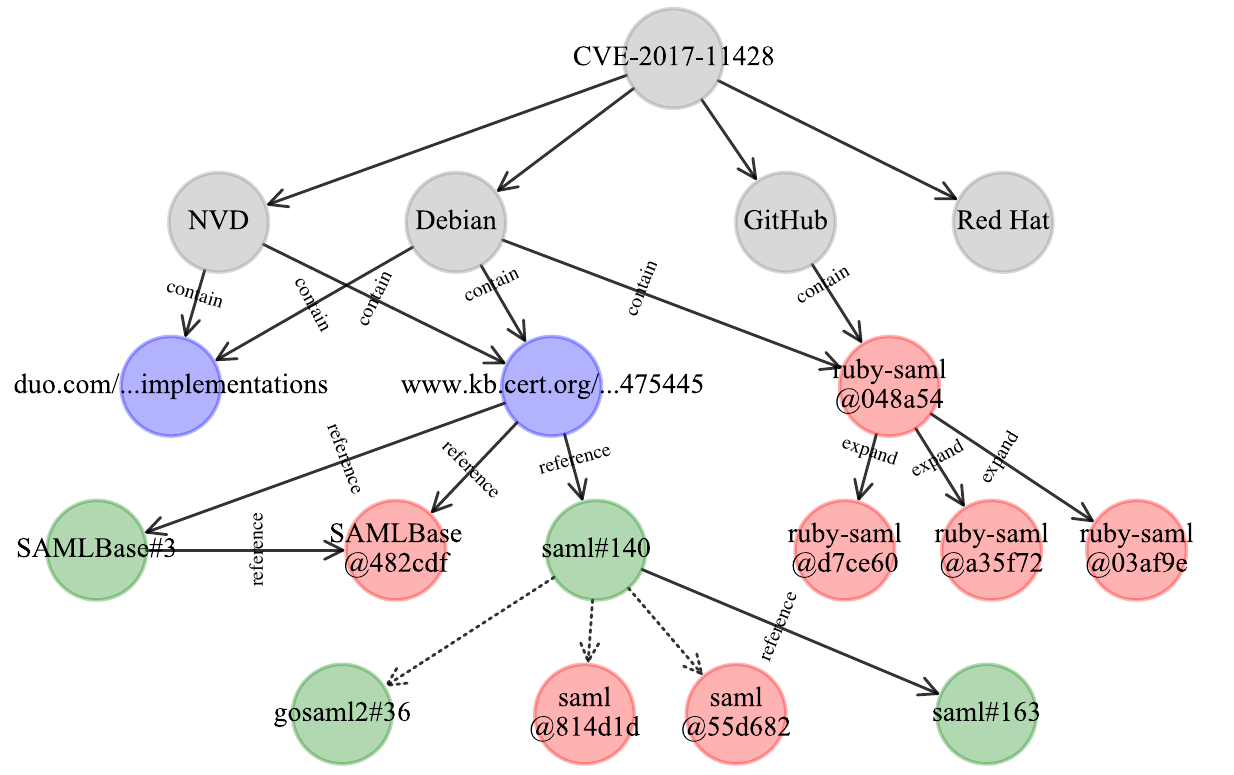}
\vspace{-10pt}
\caption{Reference Network for CVE-2017-11428}\label{fig:example}
\end{figure}

\begin{exmp}
Fig.~\ref{fig:example} presents the complete reference network for CVE-2017-11428. The top layer shows the root node, and the second layer shows the advisory source nodes.
\end{exmp}

Then, \tool respectively requests the advisory from NVD, Debian and Red Hat with the CVE identifier. Specifically, as NVD~provides structured data feeds~\cite{nvd-feed} of all vulnerabilities in the form~of JSON by year, \tool requests and parses the corresponding JSON file to obtain the NVD advisory. As Debian stores advisories~at~a repository~\cite{debian-repo}, \tool extracts the Debian advisory from~the~repository. As Red Hat provides WebService API~\cite{redhat-api}, \tool~uses it to retrieve the Red Hat advisory. Notice that Debian tracks all CVEs on NVD, and Red Hat tracks some of them.

\tool extracts URL references in each requested advisory and adds them as child nodes of the corresponding advisory~source node. For an NVD advisory, \tool extracts URL references in the ``references'' field, where references~to advisories and solutions~are listed. Similarly, for a Debian advisory, \tool extracts~URL~references in the ``Notes'' field. For a Red Hat advisory, \tool~uses a regular expression to extract URL references~in~the~``comments''~field, where developers discuss and record the resolution process of the vulnerability and may list references to patches.

\begin{exmp}
As shown in the third layer in Fig.~\ref{fig:example}, the NVD~advisory for CVE-2017-11428 contains two references. One is a reference to a blog that describes the technical detail of this vulnerability, and the other is a reference to a third-party advisory. The two~references are also contained by the Debian advisory which further~contains a reference to a GitHub commit ruby-saml@048a54~\cite{ruby-saml-1} which~fixes this vulnerability. Red Hat does not collect this CVE.
\end{exmp}

\tool also classifies these reference nodes into three types,~i.e., \textit{patch node},~\textit{issue node} and \textit{hybrid node}. We distinguish patch nodes~as our goal is to find patches for the CVE. We distinguish issues~nodes~because issue trackers may assign an issue identifier to the CVE,~where developers discuss its resolution and often list references to patches. Reference nodes that are not identified as patch or issue~nodes~are~regarded as hybrid nodes, which can be blogs, third-party advisories,~etc. Inspired by our patch type analysis (Sec.~\ref{sec:type}),~\tool~identifies~a~reference node as a patch node~if its URL contains~``git''~and~matches a regular expression of commit~identifier~(i.e.,~Git platform commits), or contains ``svn'' and matches~a~regular~expression~of~commit~identifier (i.e., SVN commits). \tool identifies a reference~node as an issue node if its URL contains ``/github.com/'' and ``/issues/''~(i.e.,~GitHub issues), contains ``/github.com/'' and ``/pull/''  (i.e., GitHub pull~requests), or contains~one of the keys ``bugzilla", ``jira", ``issues", ``bugs", ``tickets" and ``tracker"~and matches a regular expression of issue identifier (i.e., issues from other issue trackers).

\begin{exmp}
As shown in the third layer in Fig.~\ref{fig:example}, the two~references contained in both the NVD and Debian advisory are identified as~hybrid nodes (i.e., the two purple nodes). The reference that is only contained in the Debian advisory is successfully identified as a patch node (i.e., the red node).
\end{exmp}

\textbf{Reference Analysis.} For each of the reference nodes constructed in the previous sub-step, \tool applies the following two analyses to construct the reference network in a layered way.

If the reference node is a patch node, \tool requests~the~commit and analyzes whether it only changes test code~or~non-source~code files. If yes, this patch node is removed from the reference network as it cannot be the patch. \congyingEdit{\tool identifies~test~code changes by checking the “test” string in paths of modified~files,~and~identifies non-source code changes by checking the suffix of modified files.}

If the reference node is an issue or hybrid node, \tool first requests the URL and gets the response (i.e., HTML text). Then,~it~uses a regular expression to extract URL references in plain text, and~uses an HTML parser to get URL references in hyperlinks (i.e.,~<a>~tags). These extracted URL references are then checked in the same~way~as the previous sub-step to identify patch and issue references, which are added as child nodes of the reference node under analysis.~No more hybrid references will be added after this layer~as the~deeper we~explore the reference network, the more~noise would be introduced by hybrid references. In other words, only~the~hybrid~references directly contained in the NVD, Debian and Red Hat advisories are included in the reference network.~There~is one exception in the above analysis for URL references~to~GitHub issues or commits. GitHub issue reports often contain references~to~issues or commits from other repositories, which can introduce~noise~to~the reference network.
\congyingEdit{To this end, if the reference~node under~analysis corresponds to a GitHub issue, its extracted URL reference~that~is~not~from the same repository will not be added to the reference~network.}

\tool repeats the above two analyses on the newly-added nodes until there is no newly-added node or the depth of our reference network reaches a limit (which is \tocheck{5} by default).

\begin{exmp}
In the first iteration, \tool keeps the patch~node ruby-saml@048a54 in the third~layer in Fig.~\ref{fig:example} because it changes non-test source code files. It identifies that one~hybrid node~in~the third layer does not reference to any issue/commit, and the~other~references to two issues SAMLBase\#3~and~saml\#140 and one commit SAMLBase@482cdf. In the next iteration, it finds that~SAMLBase\#3 references to SAMLBase@482cdf, and saml\#140 references to two issues gosaml2\#36 and saml\#163 and two commits saml@814d1d and saml@55d682. However, gosaml2\#36 is not included in our reference network as it belongs to a different repository than saml\#140; and saml@814d1d and saml@55d682 are also not included because they only change test code. Notice that these not included nodes~are still shown in Fig.~\ref{fig:example} for the ease of presentation, but are connected~by dotted arrow lines. After this iteration, the depth limit is reached.
\end{exmp}

\textbf{Reference Augmentation.} Besides NVD, Debian and Red Hat which are explicit advisory sources, repository hosting platforms can~be~regarded as an implicit source because patches are often~hidden in~the commit history. Hence, in this sub-step, \tool~searches repository hosting platforms for patch commits of the CVE in order~to~further augment the reference network.

Inspired by our patch type analysis (Sec.~\ref{sec:type}), here we only search GitHub as most patches are in the type of GitHub commits. Besides, issues trackers often assign an issue identifier to the CVE.~Similarly, advisory publishers usually assign an advisory identifier to~the~CVE. For example, the vendor advisory of CVE-2019-10426 assigns~an~advisory identifier of SECURITY-1573~\cite{SECURITY-1573}, and the issue tracker assigns an issue identifier of THRIFT-4647~\cite{THRIFT-4647} to CVE-2018-11798. Hence, \tool uses a regular expression to extract~issue~and~advisory identifiers respectively from the~URL~of~issue~and~hybrid~nodes in our reference network constructed in the previous two sub-steps.

Then, \tool uses the CVE identifier and extracted issue and advisory identifiers as the key to search for commits by REST~API~\cite{github-api-1} provided by GitHub. This API returns up to 1,000 results~for~a~search. To reduce noise, for each returned commit, \tool~checks~whether its owner and repository name matches the vendor and product~name of any CPE of the CVE. CPE is a structured naming scheme~for~affected software of the CVE, which can be parsed from the JSON~file from NVD. Here we follow Dong et al.'s matching criterion~\cite{dong2019towards}~to~have the flexibility to handle the slightly different format of the same~software name; i.e., two \congyingEdit{software} names are regarded as a match if the number~of matched words is not less than the number of unmatched words.~Besides, \tool also checks whether the commit changes non-test source code files. If both checks are passed, \tool adds it as a child node of the advisory source node of GitHub.

\begin{exmp}
For CVE-2017-11428, \tool fails to extract~any issue or advisory identifier. Thus, it uses the CVE identifier~to~search for GitHub commits. The matched commit is ruby-saml@048a54,~and its owner and repository name is ``onelogin''  and ``ruby-saml''.~As~the vendor and product name in the CPE of this CVE is ``onelogin'' and ``ruby-saml'', a complete match is achieved. As this commit is already included in our reference network, \tool connects it as a child node of the new advisory source node of GitHub, as shown in Fig.~\ref{fig:example}. 
\end{exmp}

\subsection{Patch Selection}\label{sec:selection}

The second step of \tool is to select patches from our reference network for the CVE under analysis accurately and completely.~To this end, we use two heuristics, and combine their selected~patches.

\textbf{Confidence.} We directly select the patch nodes that we treat~as having high confidence of being the correct patch for the CVE~under analysis. Specifically, we consider two kinds of patch nodes~in~our reference network as having such high confidence. First, the patch nodes that are a direct child node of the advisory source~node~of~NVD are considered as having high confidence. The reason is that~NVD~is established with a strong community effort, each vulnerability~is~manually confirmed with several procedures, and~the~data~can be continuously updated after the initial vulnerability~reporting. 
Second, the patch nodes that are a direct child node of the advisory~source node~of~GitHub are considered as having high confidence.~The~reason is that the way \tool adds such patch nodes ensures~that~the commit message contains the CVE identifier, advisory identifier, or~issue identifier of the CVE and the name of its belonging owner and repository matches the vendor and product name of the CPE. 

\begin{exmp}
From the reference network for CVE-2017-11428 in Fig.~\ref{fig:example}, \tool directly selects the patch node ruby-saml@048a54 because it is a child node of the advisory source node of GitHub~and is considered as having high confidence of being the correct~patch for CVE-2017-11428. In fact, this commit is one of the correct~patches.
\end{exmp}

\textbf{Connectivity.} The confidence-based heuristic alone is often~not strong enough to locate patches accurately and completely because NVD may not contain patch references, and CVE identifier, advisory identifier~or~issue identifier of a CVE might not be contained~in~commit messages. Hence, inspired by the idea~that~a~correct~patch~would be frequently referenced during the reporting, discussion, and resolution of a CVE in various advisory~sources (i.e., the patch node would be widely connected to the root node in our reference network), we further design a connectivity-based heuristic.

Specifically, we measure the connectivity of a patch node to the root node in our reference network based on two dimensions. First, the more paths the root~node can reach a patch node, the~higher~connectivity to the root node the patch node has. Second, the shorter~the paths from the root node to a patch node, the higher connectivity~to the root node the patch node has. To combine these two~dimensions, we use Eq.~\ref{eq:connectivity} to compute the connectivity of a patch~node~to~the~root node, where $p = 1,  ..., n$ denotes one of the $n$ paths from the~root~node to the patch node, and $d_p$ denotes the length of a path $p$. Considering the high confidence of the two advisory sources of NVD~and~GitHub, the length of a path is changed by a decrease of 1 if the path originates from the advisory source node of NVD~and~GitHub.
\begin{equation}\label{eq:connectivity}
 connectivity =\sum_{p=1}^{n}   \frac{1}{2^{({d}_{p} -1)}}
\end{equation}

Based on the connectivity of each patch node to the root node, \tool selects the patch nodes with the highest connectivity.

\begin{exmp}
In Fig.~\ref{fig:example}, there are two paths from root node~to~the patch node ruby-saml@048a54. One originates from Debian~with a length of 2, and has connectivity of 0.5. The other originates~from GitHub with an original length of 2 and a changed length of 1, and has connectivity of 1. Thus, the connectivity of ruby-saml@048a54 to the root node is 1.5. Similarly, there exist four paths from~the~root node to the patch node SAMLBase@482cdf, respectively having connectivity of 0.5, 0.25, 0.25, and 0.125. Hence, the connectivity of SAMLBase@482cdf to the root node is 1.125. \tool selects~ruby-saml@048a54 as the patch as it has the highest connectivity.
\end{exmp}

\subsection{Patch Expansion}

The third step of \tool is to expand the patches selected~in~the second step by searching relevant commits across branches~of~the~same repository. It is inspired by our cardinality analysis~(Sec.~\ref{sec:cardinality})~where we find more than \tocheck{40\%} of the CVEs have a one-to-many~mapping to their patches, and these multiple patches often locate~in~one~branch of a repository (as a CVE is difficult to fix or the first patch~is~not~complete) or multiple branches of a repository (as a CVE affects multiple versions whose branches are separately maintained). For these~multiple patches, our reference network constructed in the previous~two steps often does not capture them completely. Besides, our patch~type analysis (Sec.~\ref{sec:type}) shows that most patches are in the type~of~GitHub commits. Therefore, the third step of \tool is designed as follows: for each selected patch that is in the type of a GitHub commit, \tool locates its repository, collects all branches in this repository, and searches the commits within a specific span of each branch for commits that are potentially patches for the CVE under analysis.

Specifically, for a selected patch that is in the type of a GitHub~commit, \tool uses a regular expression to extract the owner~and~repository information from the patch URL,
\congyingEdit{owing~to~the~well-structured commit URL in GitHub.}
Based on the owner and repository~data, \tool retrieves all branches in the repository by GitHub's~REST API~\cite{github-api-2}. Then, for each branch, \tool retrieves the commits created before and after the selected patch within a specific span~(which is \tocheck{30 days} by default) by GitHub's REST API~\cite{github-api-3}. We~do~not~retrieve all the commits for balancing performance and accuracy.~Then, for each retrieved commit, \tool uses the following two criteria~to~determine whether the commit is the patch for the CVE under analysis: i) the commit message of the retrieved commit is the same as,~contains, or is contained by the commit message of the selected patch; or ii) the commit message of the retrieved commit contains the CVE identifier, advisory identifier or issue identifier.~If~a~retrieved~commit satisfies one of the two criteria, \tool also adds~such~expanded patches as child nodes of the selected patch.

Finally, \tool returns the selected patches in the second~step and the expanded patches in the third step as the patches for the CVE under analysis.
\congyingEdit{Besides, our reference network is also returned for the ease of confirming returned patches.}

\begin{exmp}
For the selected patch ruby-saml@048a54  (locating on the master branch) for CVE-2017-11428, \tool expands~it~by~finding three commits ruby-saml@d7ce60~\cite{ruby-saml-2}, ruby-saml@a35f72~\cite{ruby-saml-3} and ruby-saml@03af9e~\cite{ruby-saml-4} which have the same commit message to ruby-saml@048a54 but respectively locate on branches 0.8.3 -- 0.8.17, v0.9.3 and v1.6.2. As shown in Fig.~\ref{fig:example}, \tool adds them as child nodes of ruby-saml@048a54. Notice that these four patches are all correct and involve different code changes. The two vulnerability databases in Sec.~\ref{sec:study} only report the patch ruby-saml@048a54.
\end{exmp}




\section{Evaluation}

\textbf{Research Questions.} We design our evaluation to answer~the~following \congyingEdit{four} research questions.

\begin{itemize}[leftmargin=*]
\item \textbf{RQ6 Effectiveness Evaluation:} how is the effectiveness~of~\tool in~tracking patches, compared to existing heuristic-based approaches and \congyingEdit{two industrial} vulnerability databases? (Sec.~\ref{sec:accuracy-evaluation})
\item \textbf{RQ7 Ablation Analysis:} how is the contribution of each component in \tool to its achieved effectiveness? (Sec.~\ref{sec:ablation})
\item \textbf{RQ8 Generality Evaluation:} how is the generality of \tool to \congyingEdit{OSS} vulnerabilities beyond our depth dataset? (Sec.~\ref{sec:generality})
\item \textbf{RQ9 Usefulness Evaluation:} how is the usefulness of \tool in practice? 
(Sec.~\ref{sec:usefulness})
\end{itemize}

We use our depth dataset to answer \textbf{RQ6} and \textbf{RQ7},~and~build~two datasets to answer \textbf{RQ8}. \congyingEdit{We conduct a user study~to~answer~\textbf{RQ9}.}

\textbf{Evaluation Metrics.} We use four metrics to measure the effectiveness of patch tracking in \textbf{RQ6}, \textbf{RQ7} and \textbf{RQ8}. The first~metric is the number of CVEs whose patches are not found~by~a~patch~tracking approach. It measures \textit{patch coverage} by only considering~whether patches are found or not. The other metrics are precision,~recall and F1-score (i.e., the same metrics in Sec.~\ref{sec:accuracy}), which measure~\textit{patch~accuracy} for those CVEs whose patches are found by a patch~tracking approach. In \textbf{RQ9}, we use patch accuracy and the time consumed~by~users with/without the help of \tool in finding~patches.

\subsection{Effectiveness Evaluation (RQ6)}\label{sec:accuracy-evaluation}

\begin{table*}[!t]
\centering
\small
\caption{Effectiveness of Existing Heuristic-Based Approaches}\label{table:heuristic}
\vspace{-10pt}
\begin{tabular}{|c|c|*{1}{C{5.1em}}*{3}{C{2.4em}}|*{1}{C{5.1em}}*{3}{C{2.4em}}|*{1}{C{5.1em}}*{3}{C{2.4em}}|}
\noalign{\hrule height 1pt}
\multirow{2}{*}{Cardinality} & \multirow{2}{*}{Number} &  \multicolumn{4}{c|}{Searching NVD References} & \multicolumn{4}{c|}{Searching GitHub Commit History} & \multicolumn{4}{c|}{Searching NVD and GitHub} \\\cline{3-14}
& & Not Found & Pre. & Rec. & F1 & Not Found & Pre. & Rec. & F1 & Not Found &  Pre. & Rec. & F1 \\
\noalign{\hrule height 1pt}
1:1 (SP) & 567 &	285 (50.3\%) & 0.973 & 0.986 & 0.977 &	472 (83.2\%) & 0.416 & 0.642 & 0.471 &	222 (39.2\%) & 0.839 & 0.930 & 0.864 \\
1:$i$ (MEP) &195 &	125 (64.1\%) & 0.932 & 0.925 & 0.921 &	162 (83.1\%) & 0.472 & 0.490 & 0.452 &	104 (53.3\%) & 0.821 & 0.867 & 0.820 \\
1:$n$ (MP) & 101 &	68 (67.3\%) & 0.980 & 0.552 & 0.683 &	73 (72.3\%) & 0.536 & 0.445 & 0.461 &	52 (51.5\%) & 0.779 & 0.605 & 0.647 \\
1:$n$ (MB) & 372 &	244 (65.6\%) & 0.979 & 0.416 & 0.546 &	246 (66.1\%) & 0.445 & 0.236 & 0.284 &	171 (46.0\%) & 0.704 & 0.393 & 0.465 \\
1:$n$ (MR) & 60 &	46 (76.7\%) & 1.000 & 0.708 & 0.794 &	37 (61.7\%) & 0.627 & 0.345 & 0.413 &	27 (45.0\%) & 0.801 & 0.539 & 0.604 \\\hline
Total & 1,295 &	    768 (59.3\%) & 0.970 & 0.805 & 0.842 &	990 (76.4\%) & 0.461 & 0.417 & 0.386 &	576 (44.5\%) & 0.793 & 0.732 & 0.720 \\
\noalign{\hrule height 1pt}
\end{tabular}
\end{table*}


\textbf{Comparison to Heuristic-Based Approaches.} We pick two~widely used heuristics: i) searching NVD references of a CVE for commits (e.g., \cite{duan2019automating, li2016vulpecker, li2018vuldeepecker}) and ii) searching GitHub commit history~for~commits containing the~identifier of a CVE in commit messages (e.g.,~\cite{you2017semfuzz, Wang2020empirical}). The first heuristic can be used to approximate the quality~of~hidden patches in NVD. In fact, we manually track the hidden~patches in NVD, which achieves similar effectiveness results to this heuristic. 
The second~heuristic is usually used for searching patches~for a known OSS, we adapt it to search patches for a CVE~by~further checking whether~the~owner~and~repository match the vendor~and~product~in CPE (i.e.,~our~strategy~in~reference augmentation).~We~also investigate a third heuristic~that~combines the results of the above two heuristics. Table~\ref{table:heuristic} and \ref{table:contribution} respectively present~the~effectiveness results of the three heuristics and \tool (and its variants, which will be discussed in Sec.~\ref{sec:ablation}). 

On one hand, for patch coverage, all~the~three~heuristics~fail~to~find any patch (i.e., return nothing) for a very large part (i.e., \tocheck{59.3\%,~76.4\% and 44.5\%}) of the CVEs across all mapping cardinalities. Differently, \tool~fails to find a patch for only \tocheck{12.0\%} of the CVEs.~On~the~other hand, for patch accuracy on the CVEs whose patches are found,~the first heuristic achieves a high patch precision due to the high confidence of NVD references, but a low patch recall~on~the~CVEs~with~one-to-many mappings; the second heuristic has both a low~patch~precision and a low patch recall; and the third heuristic achieves a patch precision and a patch recall between the first and second heuristic. \tool has~a lower patch precision, a higher patch recall~(a~significantly higher patch recall on CVEs belonging~to \textit{MP} and \textit{MB}), and~a comparable F1-score than~the first heuristic. 
We believe~it~is~acceptable because \tool finds patches for \tocheck{116.3\%}~more CVEs for which the first heuristic fails to find any patch. Besides, \tool~improves the second and third~heuristic in F1-score by \tocheck{116.8\%} and \tocheck{16.3\%}.

\begin{tcolorbox}[size=title, opacityfill=0.15]
Existing heuristics fail to find any patch for a very large~part~of vulnerabilities, while \tool finds patches for \tocheck{58.6\%} to \tocheck{273.8\%} more vulnerabilities than~them. For the vulnerabilities whose patches are found, \tool has either~a comparable F1-score~or a higher F1-score by~up~to \tocheck{116.8\%} than existing heuristics.
\end{tcolorbox}


\begin{table*}[!t]
\centering
\small
\caption{Contribution of Each Component in \tool}\label{table:contribution}
\vspace{-10pt}
\begin{tabular}{|c|c|*{1}{C{5.1em}}*{3}{C{2.4em}}|*{1}{C{5.1em}}*{3}{C{2.4em}}|*{1}{C{5.1em}}*{3}{C{2.4em}}|}
\noalign{\hrule height 1pt}
\multirow{2}{*}{Cardinality} & \multirow{2}{*}{Number} &  \multicolumn{4}{c|}{ \tool } & \multicolumn{4}{c|}{$v_1^1$: \tool w/o NVD} & \multicolumn{4}{c|}{$v_1^2$: \tool w/o Debian} \\\cline{3-14}
& & Not Found & Pre. & Rec. & F1 & Not Found & Pre. & Rec. & F1 & Not Found &  Pre. & Rec. & F1 \\
\noalign{\hrule height 1pt}
1:1 (SP) & 567 &	102 (18.0\%) & 0.860 & 0.951 & 0.881 &	286 (50.4\%) & 0.820 & 0.936 & 0.846 &	110 (19.4\%) & 0.847 & 0.943 & 0.869 \\
1:$i$ (MEP) &195 &	6 (3.1\%) & 0.886 & 0.918 & 0.888 &	    79 (40.5\%) & 0.882 & 0.935 & 0.886 &	8 (4.1\%) & 0.880 & 0.912 & 0.882 \\
1:$n$ (MP) & 101 &	20 (19.8\%) & 0.872 & 0.741 & 0.761 &	41 (40.6\%) & 0.881 & 0.728 & 0.766 &	22 (21.8\%) & 0.851 & 0.716 & 0.739 \\
1:$n$ (MB) & 372 &	23 (6.2\%) & 0.861 & 0.788 & 0.795 &	84 (22.6\%) & 0.876 & 0.780 & 0.800 &	28 (7.5\%) & 0.838 & 0.760 & 0.771 \\
1:$n$ (MR) & 60 &	4 (6.7\%) & 0.831 & 0.620 & 0.659 &	    8 (13.3\%) & 0.848 & 0.551 & 0.624 &	5 (8.3\%) & 0.819 & 0.613 & 0.651 \\\hline
Total & 1,295 &	    155 (12.0\%) & 0.864 & 0.864 & 0.837 &	498 (38.5\%) & 0.856 & 0.839 & 0.815 &	173 (13.4\%) & 0.848 & 0.849 & 0.821 \\
\noalign{\hrule height 1pt}

\multirow{2}{*}{Cardinality} & \multirow{2}{*}{Number} &  \multicolumn{4}{c|}{$v_1^3$: \tool w/o Red Hat} & \multicolumn{4}{c|}{$v_1^4$: \tool w/o GitHub} & \multicolumn{4}{c|}{$v_1^5$: \congyingEdit{\tool w/o Network}} \\\cline{3-14}
& & Not Found & Pre. & Rec. & F1 & Not Found & Pre. & Rec. & F1 & Not Found &  Pre. & Rec. & F1 \\
\noalign{\hrule height 1pt}
1:1 (SP) & 567 &	113 (19.9\%) & 0.853 & 0.943 & 0.874 &	149 (26.3\%) & 0.898 & 0.943 & 0.908 &	177 (31.2\%) & 0.910 & 0.972 & 0.925 \\
1:$i$ (MEP) &195 &	7 (3.6\%) & 0.883 & 0.918 & 0.886 &	    19 (9.7\%) & 0.887 & 0.921 & 0.892 &	78 (40.0\%) & 0.956 & 0.959 & 0.941 \\
1:$n$ (MP) & 101 &	21 (20.8\%) & 0.880 & 0.736 & 0.760 &	28 (27.7\%) & 0.873 & 0.690 & 0.726 &	40 (39.6\%) & 0.943 & 0.669 & 0.743 \\
1:$n$ (MB) & 372 &	35 (9.4\%) & 0.844 & 0.761 & 0.767 &	39 (10.5\%) & 0.874 & 0.752 & 0.773 &	109 (29.3\%) & 0.908 & 0.575 & 0.659 \\
1:$n$ (MR) & 60 &	4 (6.7\%) & 0.738 & 0.640 & 0.618 &	    7 (11.7\%) & 0.816 & 0.545 & 0.604 &	10 (16.7\%) & 0.920 & 0.641 & 0.712 \\\hline
Total & 1,295 &	    180 (13.9\%) & 0.851 & 0.853 & 0.823 &	242 (18.7\%) & 0.883 & 0.841 & 0.835 &	414 (32.0\%) & 0.918 & 0.812 & 0.823 \\
\noalign{\hrule height 1pt}

\multirow{2}{*}{Cardinality} & \multirow{2}{*}{Number} &  \multicolumn{4}{c|}{$v_2^1$: \congyingEdit{\tool w/o Selection}} & \multicolumn{4}{c|}{$v_2^2$: \tool w/o Connectivity} & \multicolumn{4}{c|}{$v_2^3$: \tool w/o Confidence} \\\cline{3-14}
& & Not Found & Pre. & Rec. & F1 & Not Found & Pre. & Rec. & F1 & Not Found &  Pre. & Rec. & F1 \\
\noalign{\hrule height 1pt}
1:1 (SP) & 567 &	102 (18.0\%) & 0.632 & 0.961 & 0.680 &	245 (43.2\%) & 0.892 & 0.978 & 0.913 &	102 (18.0\%) & 0.860 & 0.942 & 0.879 \\
1:$i$ (MEP) &195 &	6 (3.1\%) & 0.622 & 0.976 & 0.682 &	    111 (56.9\%) & 0.929 & 0.939 & 0.915 &	6 (3.1\%) & 0.888 & 0.913 & 0.889 \\
1:$n$ (MP) & 101 &	20 (19.8\%) & 0.615 & 0.933 & 0.656 &	56 (55.4\%) & 0.953 & 0.685 & 0.764 &	20 (19.8\%) & 0.880 & 0.722 & 0.751 \\
1:$n$ (MB) & 372 &	23 (6.2\%) & 0.616 & 0.903 & 0.657 &	191 (51.3\%) & 0.927 & 0.787 & 0.821 &	23 (6.2\%) & 0.871 & 0.765 & 0.784 \\
1:$n$ (MR) & 60 &	4 (6.7\%) & 0.368 & 0.891 & 0.394 &	    27 (45.0\%) & 0.885 & 0.722 & 0.772 &	4 (6.7\%) & 0.849 & 0.462 & 0.550 \\\hline
Total & 1,295 &	155 (12.0\%) & 0.611 & 0.940 & 0.658 &	    630 (48.6\%) & 0.910 & 0.889 & 0.871 &	155 (12.0\%) & 0.869 & 0.844 & 0.826 \\

\noalign{\hrule height 1pt}

\multirow{2}{*}{Cardinality} & \multirow{2}{*}{Number} &  \multicolumn{4}{c|}{$v_2^4$: \tool with Path Length} & \multicolumn{4}{c|}{$v_2^5$: \tool with Path Number} & \multicolumn{4}{c|}{$v_3$: \tool w/o Expansion} \\\cline{3-14}
& & Not Found & Pre. & Rec. & F1 & Not Found & Pre. & Rec. & F1 & Not Found &  Pre. & Rec. & F1 \\
\noalign{\hrule height 1pt}
1:1 (SP) & 567 &	102 (18.0\%) & 0.833 & 0.957 & 0.859 &	102 (18.0\%) & 0.805 & 0.951 & 0.837 &  102 (18.0\%) & 0.871 & 0.948 & 0.889\\
1:$i$ (MEP) &195 &	6 (3.1\%) & 0.848 & 0.945 & 0.867 &	    6 (3.1\%) & 0.849 & 0.920 & 0.858 &     6 (3.1\%) & 0.910 & 0.914 & 0.902\\
1:$n$ (MP) & 101 &	20 (19.8\%) & 0.849 & 0.760 & 0.742 &	20 (19.8\%) & 0.801 & 0.756 & 0.726 &   20 (19.8\%) & 0.873 & 0.696 & 0.732\\
1:$n$ (MB) & 372 &	23 (6.2\%) & 0.830 & 0.798 & 0.770 &	23 (6.2\%) & 0.833 & 0.811 & 0.791 &    23 (6.2\%) & 0.860 & 0.506 & 0.590\\
1:$n$ (MR) & 60 &	4 (6.7\%) & 0.652 & 0.747 & 0.590 &	    4 (6.7\%) & 0.789 & 0.630 & 0.644 &     4 (6.7\%) & 0.847 & 0.567 & 0.629\\\hline
Total & 1,295 &	155 (12.0\%) & 0.827 & 0.882 & 0.812 &	    155 (12.0\%) & 0.819 & 0.873 & 0.809 &  155 (12.0\%) & 0.873 & 0.771 & 0.776\\
\noalign{\hrule height 1pt}
\end{tabular}
\end{table*}

\textbf{Comparison to Vulnerability Databases.} We are~not~aware~of how much manual effort or what automated approach is involved in the construction of industrial vulnerability databases. Some~of~them claimed that they collect patches manually or semi-automatically. Therefore, the goal~of~our~comparison to \congyingEdit{industrial} databases~is~not~to demonstrate~the superiority or inferiority of \tool over industrial databases, but~to~assess~the~level~of~effectiveness \tool~can~achieve and the worthiness of \tool,~and~explore whether \tool~can~potentially improve or complement~existing industrial databases.

As shown in Table~\ref{table:accuracy} and \ref{table:contribution}, \tool finds patches~for \tocheck{12.0\%} \congyingEdit{fewer} CVEs than $DB_A$ and $DB_B$. This is determined by the way we construct this used depth dataset (i.e., our depth dataset includes vulnerabilities whose patches are provided by $DB_A$ and $DB_B$). We will demonstrate in Sec.~\ref{sec:generality} how \tool works on vulnerabilities whose patches are not provided by $DB_A$ and $DB_B$. 

On the CVEs whose patches~are~found,~\tool has~a~\tocheck{6.4\%}~and \tocheck{5.8\%} lower patch precision than $DB_A$ and $DB_B$. This might~be~potentially due to the manual effort involved in industrial database~construction. Besides, \tool has a \tocheck{15.5\% and 18.4\%} higher patch~recall than $DB_A$ and $DB_B$ (especially on CVEs belonging to one-to-many mappings), resulting in a \tocheck{5.5\% and 8.6\%} higher F1-score~than $DB_A$ and $DB_B$. These results indicate that  \tool is worthwhile~with~a~significantly higher patch recall at the price of a moderately lower~patch precision. \tool has the merit to complement industrial databases by reducing manual effort and tracking patches completely.

\begin{tcolorbox}[size=title, opacityfill=0.15]
\tool achieves a \tocheck{15.5\% to 18.4\%} higher patch recall and~a~\tocheck{5.5\% to 8.6\%} higher F1-score than the two industrial vulnerability databases, while sacrificing up to \tocheck{6.4\%} lower patch precision. \tool can complement industrial vulnerability databases by reducing manual effort and tracking patches completely.
\end{tcolorbox}


\textbf{False Negative Analysis.} We manually analyze the CVEs for which \tool finds no patch or misses some of the patches,~and summarize five main reasons. First, for some old CVEs, the references contained in NVD, Debian and Red Hat are limited,~and~some of them even become invalid. As a result, \tool fails to construct a complete reference network. 
Second, some key references (e.g., issue reports) about a CVE are missing from NVD, Debian~and~Red~Hat. As a result, \tool fails to be directed to the correct patch. For example, for CVE-2018-14642, its issue report~\cite{UNDERTOW-1430}~is~not~contained~in~any of the three advisory sources. However, following~the~issue report, we could find the patch~\cite{undertow}. Third, the commit message~of~a patch has semantic similarity to the CVE description, but does~not contain the CVE identifier.~Hence, our reference augmentation fails~to~catch it. For example, for CVE-2019-10077~\cite{CVE-2019-10077}, our patch commit~\cite{jspwiki}~fixes~it without indicating the CVE identifier. Fourth, GitHub's REST~API~for commit search returns 1,000 results, which may miss~the~correct~patch commit in our reference~augmentation. Fifth, only one~patch~with~the highest connectivity is selected in our patch selection.~Thus, correct patches for CVEs belonging to one-to-many mappings~might be missed although they are already included in our reference network.

\textbf{False Positive Analysis.} We also manually analyze the CVEs~for which \tool finds incorrect patches, and summarize two~main reasons. First, the commit that introduces the CVE is referenced during the discussion and resolution of the CVE. As a result, \tool falsely identifies it as a patch commit due to the lack of semantic~understanding of the context where the commit is referenced.~For~example, for CVE-2020-5249, the commit that introduces the CVE~\cite{go-ethereum-1} and the commit that fixes the CVE~\cite{go-ethereum-2} are referenced~in~the~same~comment of the issue report. Second, multiple CVEs and their issues~and patches are listed on the same page. As a result, patches for other CVEs might be falsely identified by \tool due to the lack~of semantic understanding. For example, for CVE-2018-15750, its patches are maintained in the release note~\cite{release-note} with CVE-2018-15751, and the release note is all referenced by NVD, Debian, and Red Hat.

\begin{tcolorbox}[size=title, opacityfill=0.15]
The five and two reasons for false negatives and false positives are respectively summarized by manual analysis, which can~be leveraged to further improve the effectiveness of \tool.
\end{tcolorbox}

\subsection{Ablation Analysis (RQ7)}\label{sec:ablation}
Table~\ref{table:contribution} presents the results of our ablation study to measure~the~contribution of various settings in \tool to its achieved effectiveness. 

\textbf{Removing an Advisory Source.} We remove one of the four~advisory sources NVD, Debian, Red Hat and GitHub from the first~step of \tool, and generate variants $v_1^1$, $v_1^2$, $v_1^3$ and $v_1^4$ of \tool. These four variants suffer~an~increase in the number of CVEs~they~fail~to~find any patch for. $v_1^1$,~$v_1^2$, $v_1^3$ and $v_1^4$ respectively find patches for \tocheck{30.1\%, 1.6\%, 2.2\% and 7.6\%} \congyingEdit{fewer} CVEs than \tool, while achieving comparable precision, recall and F1-score on  CVEs they find patches for. These results~indicate that all the four advisory sources contribute to finding patches~for more CVEs by constructing a more complete reference network, while NVD and GitHub contribute the most.

\textbf{Removing Reference Network.} We do not construct the reference network in a layered way but simply use the direct references contained in the four advisory sources (i.e., we skip reference analysis in the first step of \tool), which is the variant $v_1^5$ ~in~Table~\ref{table:contribution}. $v_1^5$ also suffers an increase in the number of CVEs~it fails to find~any patch for. $v_1^5$ finds patches for \tocheck{22.7\%}~fewer~CVEs~than \tool, while having a \tocheck{6.3\%} higher precision, a \tocheck{6.0\%} lower recall, and a comparable F1-score on CVEs it finds patches for.~These results demonstrate that patches are not always directly referenced in NVD, Debian~and Red Hat, but might be hidden in indirect references, and our reference network contributes to tracking such hidden patches at the price of an acceptable decrease in precision.

\textbf{Removing Patch Selection.} We do not select some patches~in the second step of \tool but select all the patches in our reference network, which is variant $v_2^1$ in Table~\ref{table:contribution}. $v_2^1$ significantly improves \tool in recall by \tocheck{8.8\%}, especially for CVEs belonging to one-to-many mappings, while suffering a large degradation~in precision by \tocheck{29.3\%} and in F1-score by \tocheck{21.4\%} across all cardinalities. These results indicate that our reference network indeed contains most of the correct patches, and our patch selection heuristics contribute to achieving a balance between precision and recall.

\textbf{Removing Connectivity or Confidence.} We remove one~of~the two heuristics adopted in the second step of \tool, and generate two variants $v_2^2$ and $v_2^3$. Without our connectivity-based~heuristic, $v_2^2$ finds patches for \tocheck{41.7\%} \congyingEdit{fewer} CVEs than \tool, while~achieving a \tocheck{5.3\%} higher precision, a \tocheck{2.9\%} higher recall and a \tocheck{4.1\%} higher F1-score. These results indicate that our connectivity-based heuristic contributes to finding patches~for more CVEs while introducing~acceptable noise.  Without our confidence-based heuristic, $v_2^3$ suffers~a slight decrease in recall and F1-score, especially for CVEs belonging to \textit{MR}. These results show that our confidence-based heuristic contributes to achieving a balanced accuracy across all cardinalities.

\textbf{Reducing Connectivity.} We reduce connectivity-based~heuristic by only considering path length (i.e., selecting the patch with~the shortest path to the root node) and by only considering path number (i.e., selecting the patch with the largest number of paths~to~the~root node), and respectively generate the variant $v_2^4$ and $v_2^5$. Both $v_2^4$~and $v_2^5$ suffer a \tocheck{4.3\%} and \tocheck{5.2\%} decrease in precision, a \tocheck{2.1\%} and \tocheck{1.0\%}~increase in recall, and a \tocheck{3.0\%} and \tocheck{3.3\%} decrease in F1-score. These results demonstrate that our connectivity-based~heuristic contributes to improving the precision of \tool by comprehensively considering the path length and path number of a patch to the root~node.

\textbf{Removing Patch Expansion.} We do not expand patches~in~the third step of \tool, which is variant $v_3$ in Table~\ref{table:contribution}. $v_3$ suffers~degradation~in~recall~and~F1-score by \tocheck{10.8\%}  and \tocheck{7.3\%}, especially for CVEs with one-to-many~mappings. These results show that our patch expansion contributes to finding multiple patches more completely.

\begin{tcolorbox}[size=title, opacityfill=0.15]
Our used advisory sources, reference network, patch selection, and patch expansion all contribute positively to the achieved effectiveness of \tool in tracking patches.
\end{tcolorbox}


\subsection{Generality Evaluation (RQ8)}\label{sec:generality}

\begin{table*}[!t]
\centering
\small
\caption{Generality of \tool over Two New Datasets}\label{table:generality}
\vspace{-10pt}
\begin{tabular}{|c|*{1}{C{5.1em}}|*{3}{C{2.6em}}|*{1}{C{5.1em}}*{3}{C{2.6em}}|*{1}{C{5.1em}}*{3}{C{2.6em}}|}
\noalign{\hrule height 1pt}
\multirow{2}{*}{Dataset} & \multirow{2}{*}{Number} &  \multicolumn{3}{c|}{\tool} & \multicolumn{4}{c|}{$DB_A$} & \multicolumn{4}{c|}{$DB_B$} \\\cline{3-13}
& & Pre. & Rec. & F1 & Not Found & Pre. & Rec. & F1  & Not Found &  Pre. & Rec. & F1 \\
\noalign{\hrule height 1pt}
First Dataset & 91 & 0.823 & 0.845 & 0.784  & 29 (31.9\%) &  0.935 & 0.827 & 0.858  & 62 (68.1\%) & 0.885 & 0.664 & 0.725\\\hline
Second Dataset & 89 & 0.888 & 0.899 & 0.867 & --& -- & -- & -- & -- & -- & -- & -- \\\hline
\noalign{\hrule height 1pt}
\end{tabular}
\end{table*}

To evaluate the generality of \tool (i.e., whether \tool~is~overfitted to our depth dataset), we collect~two new vulnerability~datasets, and run \tool against them. The first dataset includes the~CVEs~for which only one of the \congyingEdit{two} industrial vulnerability databases~reports their patches (Fig.~\ref{fig:rq1-cves-with-patches}), which has \tocheck{3,185} CVEs. The second~dataset~includes the CVEs for which none of~the~two industrial vulnerability databases reports their patches (Fig.~\ref{fig:intersection}), which has \tocheck{5,468}~CVEs.

\tool finds patches for \tocheck{2,155 (67.7\%)} of the \tocheck{3,185} CVEs in the first dataset, and the two industrial vulnerability databases~$DB_A$~and $DB_B$ report patches for \tocheck{2,190 (68.8\%)} and \tocheck{995 (31.2\%)} CVEs. Of the \tocheck{2,155} CVEs \tool finds patches for, $DB_A$ and ~$DB_B$ report~patches for \tocheck{1,455} and \tocheck{700} CVEs. In addition, \tool finds patches for \tocheck{2,816 (51.5\%)} of the \tocheck{5,468} CVEs in the second~dataset, where $DB_A$~and~$DB_B$ report no patch. These results~indicate~that \tool complements industrial vulnerability databases by tracking patches~for~CVEs whose patches are not provided by industrial vulnerability databases.

Then, we respectively sample \tocheck{100} CVEs \tool~finds~patches~for from the first and second datasets, and manually find their patches~in the same procedure in Sec.~\ref{sec:preparation} to measure the accuracy~of~\tool. The results are reported in Table~\ref{table:generality}. Our manual analysis has~\tocheck{9}~and~\tocheck{11} uncertain CVEs due to limited disclosed information. Of the \tocheck{91} CVEs in the first dataset, \tool has an F1-score of \tocheck{0.784}, while~$DB_A$~provides patches for 62 (68.1\%) CVEs with a higher F1-score and $DB_B$ provides patches for 29 (31.9\%) CVEs with a lower F1-score. Similar to the results in Sec.~\ref{sec:accuracy-evaluation}, industrial vulnerability databases achieve a higher precision but a lower recall. Of the \tocheck{89} CVEs in the second dataset, $DB_A$ and $DB_B$ report no patch, whereas \tool achieves an F1-score of \tocheck{0.867}. These results indicate that \tool can~be~generalized to vulnerabilities beyond the ones in our depth dataset.



In addition, we find that the two industrial vulnerability databases have been updated since April 7th, 2020 (i.e., our crawling date).~Thus, we investigate the generality of \tool from another perspective, i.e., how patches tracked by \tool are similar to their updates.~To this end, we update all vulnerability entries in $DB_A$ and $DB_B$~to~$DB_A^+$ and $DB_B^+$ at March 8th, 2022. Of the CVEs whose~patches~are~found by \tool but not provided by $DB_A$ (resp. $DB_B$), 147 (resp. 669)~CVEs' patches are newly provided by~$DB_A^+$ (resp. $DB_B^+$). For 56 (38.1\%) (resp. 405 (60.5\%)) of the 147 (resp.~669) CVEs, \tool finds the~same~patches to the patches provided by $DB_A^+$ (resp. $DB_B^+$). For 37 (25.2\%) (resp. 199 (29.7\%)) of the 147 (resp.~669) CVEs, the patches provided~by~$DB_A^+$ (resp. $DB_B^+$) are included in the~patches tracked by \tool. These results indicate that the patches found by \tool have a high chance of being accepted by industrial vulnerability databases.

\begin{tcolorbox}[size=title, opacityfill=0.15]
\tool finds patches for \tocheck{67.7\%} and \tocheck{51.5\%} of the CVEs in the two new datasets with a sampled patch precision of \tocheck{0.823}~and \tocheck{0.888} and a sampled patch recall of \tocheck{0.845} and \tocheck{0.899}.~\tool~is generalizable and complements industrial databases.
\end{tcolorbox}

\subsection{\congyingEdit{Usefulness Evaluation (RQ9)}}\label{sec:usefulness}

In practice, to ensure patch accuracy, security experts still~need~to~verify patches even when automatic tools are used to find patches.~To~evaluate the usefulness of \tool in such a usage scenario,~we~conduct a user study with 10 participants who are required~to~find~patches for 10 CVEs with and without the help of \tool. We recruit~10~participants from security laboratories in multiple universities and~high-tech companies. They are Postdocs, PhD~students, master researchers, and engineers majoring in software security. We randomly select 10 CVEs from our depth dataset as tasks. 2 CVEs belong to \textit{SP},~3~CVEs belong to \textit{MEP}, 1 CVE belongs to \textit{MP}, and 4 CVEs belong to \textit{MB}. To have a fair comparison, we divide participants into two~groups (i.e., A and B). Group A is required to complete the first five tasks~without \tool (but can use existing heuristics) and finish the remaining tasks with \tool. Group B is required to complete the first five tasks with \tool,~and finish the remaining tasks without \tool. 

\begin{table*}[!t]
\centering
\small
\caption{Comparison Results of the Time and Accuracy of 10 Tasks}\label{table:usefulness}
\vspace{-10pt}
\begin{tabular}{|c|*{1}{C{5.1em}}|*{3}{C{2.6em}}|*{1}{C{5.1em}}|*{3}{C{2.6em}}|*{1}{C{5.1em}}|*{3}{C{2.6em}}|}
\noalign{\hrule height 1pt}
\multirow{2}{*}{Approach} &  \multicolumn{4}{c|}{{All 10 Tasks}} & \multicolumn{4}{c|}{5 Single-Patch Tasks} & \multicolumn{4}{c|}{5 Multiple-Patches Tasks} \\\cline{2-13}
& Time (mins) & Pre. & Rec. & F1 & Time (mins) & Pre. & Rec. & F1 & Time (mins) &  Pre. & Rec. & F1 \\
\noalign{\hrule height 1pt}
w/o \tool & 5.66 & 0.880 & 0.677 & 0.765      & 5.60 & 0.960 & 0.960 & 0.960        & 5.72 & 0.800 & 0.393 & 0.527 \\\hline
with \tool  & 4.66 & 0.983 & 0.920 & 0.951      & 3.84 & 1.000 & 1.000 & 1.000        & 5.48 & 0.967 & 0.840 & 0.899 \\\hline
\noalign{\hrule height 1pt}
\end{tabular}
\end{table*}

Table~\ref{table:usefulness} reports the average time consumption and patch accuracy of the 10 tasks. We categorize the 5 CVEs belonging~to~\textit{SP}~and~\textit{MEP}~as \textit{Single-Patch} tasks, and categorize the 5 CVEs belonging to \textit{MP}~and~\textit{MB} as \textit{Multiple-Patches} tasks. Overall, with the assistance of \tool,~the participants save time by \tocheck{17.7\%} for each task, and improve the patch accuracy in terms of precision, recall and F1-score by \tocheck{11.7\%}, \tocheck{35.9\%} and \tocheck{24.3\%}. In particular, the time saving is significant~for~the~5~\textit{Single-Patch} tasks, but not significant for the 5 \textit{Multiple-Patches}~tasks.~This~is because the reference network and patches returned by \tool for \textit{Multiple-Patches} tasks are more complex than those of \textit{Single-Patch} tasks, and participants spend more time understanding the reference network and verify patches. Moreover, the accuracy improvement~is significant for the 5~\textit{Multiple-Patches} tasks, but not significant for the 5 \textit{Single-Patch}~tasks. In that sense, \tool~is~especially~useful~for CVEs with multiple patches.

We interview each participant to get their feedback about \tool. Overall, they all appreciate the value of our reference network~as~it summarizes information from multiple sources, and the different kinds of nodes and relationships in it are helpful to localize~and~verify patches. As commented by a security engineer from a high-tech company, \textit{``the network graph, as a chain of evidence, is helpful~to~localize and verify patches"}. Moreover, they suggest including more~information for the nodes (e.g., commit message and code differences) instead of a clickable link for the ease of manual review, and also suggest adding more advisory sources. It is worth mentioning that \tool has been deployed to two high-tech companies~that~participate this user study. Unfortunately, we cannot~disclose its usage~statistics due to confidentiality agreement.

\begin{tcolorbox}[size=title, opacityfill=0.15]
\tool is useful in practice for security experts to localize patches more accurately and quickly.
\end{tcolorbox}

\subsection{Discussion}

\textbf{Limitations.} First, 
\tool only uses NVD, Debian and Red Hat~as the advisory sources in the step of advisory analysis. However,~it~is designed to easily leverage other sources~(e.g., SecurityFocus~\cite{SecurityFocus}) thanks~to~our~lightweight reference analysis. Second, as indicated~by our false negative and false positive analysis, the lack of semantic analysis of patches and vulnerabilities hurts the accuracy~of~\tool. We plan to utilize~semantics~in vulnerabilities~(e.g., descriptions)~and patches (e.g, changed code~and~the~context of patch references) to enhance patch selection and expansion. Third, \tool depends~on~the quality of the existing references in advisory sources. If there~are~not many high-quality references, \tool might be less effective. Thus, we use multiple advisory sources to reduce this possibility.

\textbf{Significance.} \tool can benefit security community, academia, and industry by enabling automated~patch tracking. For security~community,~\tool can~notify~NVD for missing~or incomplete~patches for CVEs~to enhance CVE information quality and accelerate entry update, and thus benefit the audience~of~NVD. For~academia,~\tool can enable data-driven security analysis~(e.g., learning-based vulnerability detection \cite{li2018vuldeepecker, zhou2019devign}) and empirical studies~by~providing~large-scale patches. 
\congyingEdit{For industry, \tool can assist security engineers~in enhancing patch coverage and accuracy~of industrial vulnerability databases,} and hence improve~the~accuracy of software composition analysis (i.e., determining whether vulnerabilities in patched~methods in a used OSS are reachable in an application).





\section{Related Work}


\textbf{CVE Information Quality.} 
Nguyen and Massacci~\cite{nguyen2013reliability} uncover~the unreliability of the vulnerable version data~in~NVD. To improve~its~reliability, Nguyen et al.~\cite{nguyen2016automatic}~and Dashevskyi et al.~\cite{dashevskyi2018screening} develop~tools~to determine whether older versions~are~affected~by a newly disclosed vulnerability. 
Dong~et~al. \cite{dong2019towards} identify vulnerable software~names~and versions from vulnerability reports,~and~find~that~vulnerability~data-bases miss truly vulnerable versions or falsely include non-vulnerable versions. Chen~et~al.~\cite{chen2020automated} identify open-source libraries affected~by~a vulnerability. Chaparro et al.~\cite{chaparro2017detecting} detect the absence of reproduction steps~ and expected behavior in vulnerability descriptions. Mu et al.~\cite{mu2018understanding} show the prevalence of missing reproduction information in vulnerability reports. Jo et al.~\cite{jo2020gapfinder} identify semantic inconsistencies within the cybersecurity domain. \tocheck{These works are focused~on~different aspects of vulnerability information. Following this direction, our work is focused~on~the patch~of a vulnerability.} 

A closely related work is from Tan et al.~\cite{Tan2021locating}. They use a learning-to-rank algorithm to rank commits in a repository so that patch~commits to a CVE are ranked in top positions. However, they~make~two~assumptions: i) the repository of the affected software~of~a~CVE~is~known, which is  not practical and requires manual efforts; and ii) a CVE has a one-to-one mapping to its patches, which~does~not~always hold (Sec.~\ref{sec:cardinality}). Instead,~\tool has no such assumptions. 

\textbf{Patch Analysis.}  There are many patch analysis tasks to improve security, e.g., patch generation and deployment~\cite{mulliner2013patchdroid, duan2019automating, xu2020automatic}, patch presence testing~\cite{zhang2018precise, jiang2020pdiff, dai2020bscout} and secret patch identification~\cite{xu2017spain, zhou2017automated, sabetta2018practical, chen2020machine}. Datasets of security patches have~been~built~for Java~\cite{ponta2019manually}, C/C++~\cite{fan2020ac} and specific open-source projects~\cite{jimenez2018enabling}. With such datasets, empirical studies have been conducted to characterize vulnerabilities and their patches~\cite{zaman2011security, li2017large, liu2020large, antal2020exploring}. In these~works, patches~are~identified by manual efforts \cite{xu2020automatic, jiang2020pdiff, dai2020bscout, xu2017spain, zhou2017automated, sabetta2018practical, chen2020machine, ponta2019manually,zaman2011security} or by heuristic rules like looking~for commits in CVE references~\cite{duan2019automating, fan2020ac, jimenez2018enabling, li2017large, liu2020large} and searching~for~CVE identifiers in commits~\cite{fan2020ac, jimenez2018enabling, antal2020exploring}.~\tocheck{Such heuristics only search direct references, but patches can be hidden in indirect references. \tool addresses it by a reference network.}

\textbf{Patch Applications.} Patches can be leveraged to enable various security applications,~e.g.,~generating exploits based on patches~\cite{brumley2008automatic,  you2017semfuzz}, conducting software composition analysis to determine whether vulnerabilities in a library are reachable through which call paths~\cite{pashchenko2018vulnerable, ponta2020detection, pashchenko2020vuln4real, Wang2020empirical}, and detecting vulnerabilities by learning vulnerability features~\cite{li2016vulpecker, li2018vuldeepecker, zhou2019devign, jimenez2019importance},~by matching vulnerability signatures~\cite{jang2012redebug, kim2017vuddy} and by matching both~vulnerability and patch signatures~\cite{xu2020patch, xiao2020mvp, cui2020vuldetector}. Similar to those patch analysis works, \congyingEdit{the mappings between CVEs and their patches} in these works are mostly identified by manual efforts \cite{pashchenko2018vulnerable, ponta2020detection, pashchenko2020vuln4real, xiao2020mvp} and heuristics rules~\cite{li2016vulpecker, li2018vuldeepecker, you2017semfuzz, Wang2020empirical, jimenez2019importance},~or directly taken from security advisories~that establish the mapping between CVEs and patches for specific projects~\cite{jang2012redebug, kim2017vuddy, xu2020patch}. 

\section{Conclusions}

We have conducted an empirical study to understand the quality and characteristics~of~patches for OSS vulnerabilities~in two~industrial vulnerability databases. 
We have proposed \tool to track patches~for \congyingEdit{OSS} vulnerabilities. Our extensive evaluation has demonstrated~the effectiveness, generality and usefulness of \tool. We have released the code and data at \url{https://patch-tracer.github.io}. 



\begin{acks}
This work was supported by the National Natural Science Foundation of China (Grant No. 61972098).
\end{acks}

\balance 

\bibliographystyle{ACM-Reference-Format}
\bibliography{src/reference}


\end{document}


\title{Supplementary Material for ``Tracking Patches for Open Source Software Vulnerabilities''}

\author{Congying Xu}
\affiliation{
\department{School of Computer Science and Shanghai Key Laboratory of Data Science}
\institution{Fudan University}
\city{Shanghai}
\country{China}
}

\author{Bihuan Chen}
\authornote{Bihuan Chen is the corresponding author.}
\affiliation{
\department{School of Computer Science and Shanghai Key Laboratory of Data Science}
\institution{Fudan University}
\city{Shanghai}
\country{China}
}

\author{Chenhao Lu}
\affiliation{
\department{School of Computer Science and Shanghai Key Laboratory of Data Science}
\institution{Fudan University}
\city{Shanghai}
\country{China}
}

\author{Kaifeng Huang}
\affiliation{
\department{School of Computer Science and Shanghai Key Laboratory of Data Science}
\institution{Fudan University}
\city{Shanghai}
\country{China}
}

\author{Xin Peng}
\affiliation{
\department{School of Computer Science and Shanghai Key Laboratory of Data Science}
\institution{Fudan University}
\city{Shanghai}
\country{China}
}

\author{Yang Liu}
\affiliation{
\department{School of Computer Science and Engineering}
\institution{Nanyang Technological University}
\country{Singapore}
}

\keywords{}

\maketitle

\section{Dataset Analysis}

\begin{figure*}[!t]
\centering
\begin{subfigure}[b]{0.5\textwidth}
\centering
\includegraphics[scale=0.45]{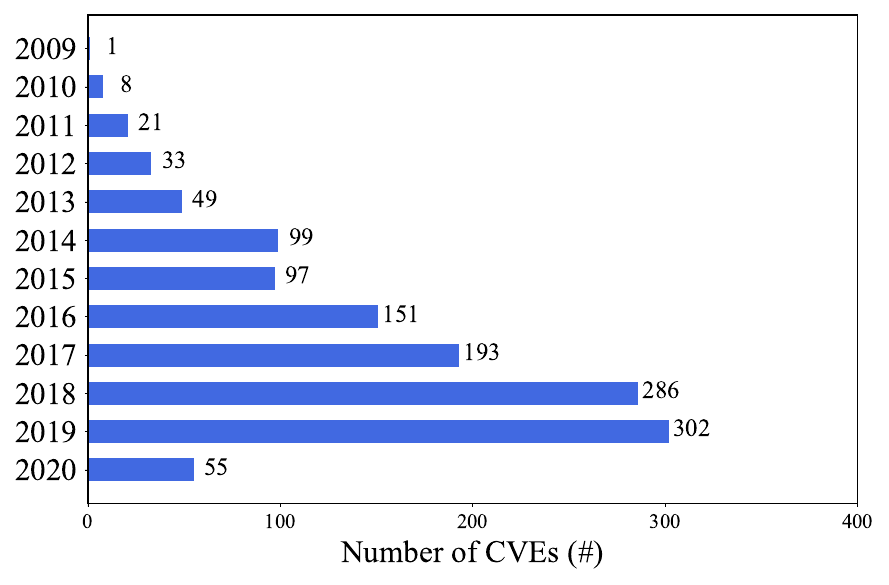}
\vspace{-5pt}
\caption{Our Depth Dataset w.r.t. Years}\label{fig:rq0-year}
\end{subfigure}
~
\begin{subfigure}[b]{0.5\textwidth}
\centering
\includegraphics[scale=0.455]{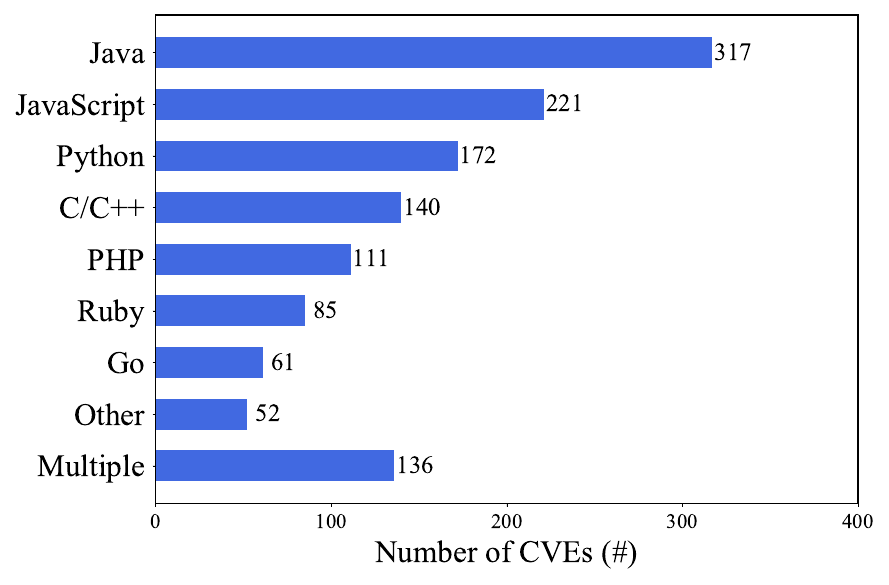}
\vspace{-5pt}
\caption{Our Depth Dataset w.r.t. Programming Languages}\label{fig:rq0-language}
\end{subfigure}
\vspace{-20pt}
\caption{Our Depth Dataset w.r.t Years and Programming Languages}\label{fig:dataset}
\end{figure*}

We analyzed the \tocheck{1,295} CVEs in our depth dataset with~respect~to~years and programming languages. We determined the~programming language~of a CVE by analyzing the changed source~files in patches.~As shown~in Fig.~\ref{fig:rq0-year}, the number of CVEs increases every year, which~is consistent with Snyk's report\footnote{https://snyk.io/wp-content/uploads/sooss\_report\_v2.pdf}. As reported in Fig.~\ref{fig:rq0-language}, these CVEs mainly cover seven programming languages, which demonstrates relatively good coverage of ecosystems. Therefore, we believe that our depth dataset is representative of \congyingEdit{OSS} vulnerabilities.


\section{Hidden Patches in NVD}

In NVD, patches for a CVE might be listed in the references,~and~references might be tagged with ``\textit{Patch}''. Therefore, we manually find such hidden patches in NVD for our depth dataset. Specifically,~we look for commit references in NVD's references and manually confirm whether they are correct patches. We also look into~the~reference tagged with ``\textit{Patch}'' to find patches that may be referenced~by the tagged reference. In this way, we find patches for 540 of the 1,295 CVEs in our depth dataset, but fail to find any patch for 755 CVEs. On those 540 CVEs, our manual effort achieves a patch precision of 1.0, a patch recall of 0.816 and an F1-score of 0.862, which~is~similar to our first heuristic in \textbf{RQ6} in the submitted~paper.

It~is~worth~mentioning that we do not include patch quality~evaluation of NVD in our empirical study because i) NVD does~not~provide a ``patch'' field for CVEs, and ii) employing the above manual analysis to measure the patch quality of NVD would be unfair~because it heavily depends on the manual effort and does not truly reflect the patch quality of NVD.

\section{Accuracy Analysis.}

We analyzed the overlap between the patches identified~by~\tool (denoted as $P_{\tool}$) and our manually identified patches (denoted~as $P_{GT}$) for each CVE in our depth dataset. In particular, we classify~the overlap relationships into six categories, which are used~as~another indicator of patch accuracy. The result is reported in Table~\ref{table:distribution}, where the first column lists the six categories, the second column shows the number of CVEs belonging to each category, and the last column gives the total number of patches found by \tool.

It can be observed that \tool can find patches accurately~and~completely for 773 (59.7\%) CVEs (i.e., $P_{\tool}$ = $P_{GT}$), with an average~of 1.9 found~patches for each CVE. \tool can find patches completely but include some false positives for 128 (9.9\%) CVEs~(i.e.,~$P_{\tool} \supset P_{GT}$). In that sense, 901 (69.6\%) CVEs' patches are completely found by \tool. Besides, \tool can find patches accurately~but~have~some false negatives for 139 (10.7\%) CVEs (i.e., $P_{\tool} \subset P_{GT}$). \tool~incurs both false positives and false negatives for 27 (2.1\%) CVEs (i.e., $P_{\tool} \cap P_{GT} \not= \emptyset$), while the patches found for 73 (5.6\%) CVEs by \tool are all false positives (i.e., $P_{\tool} \cap P_{GT}$ = $\emptyset$). Notice that we analyze the reasons for false positives and false negatives in the submitted paper. These results demonstrate the capability~of~\tool~in finding patches accurately and completely.

\section{Sensitivity Analysis}

\begin{figure*}[!t]
\centering
\begin{subfigure}[b]{0.5\textwidth}
\centering
\includegraphics[scale=0.55]{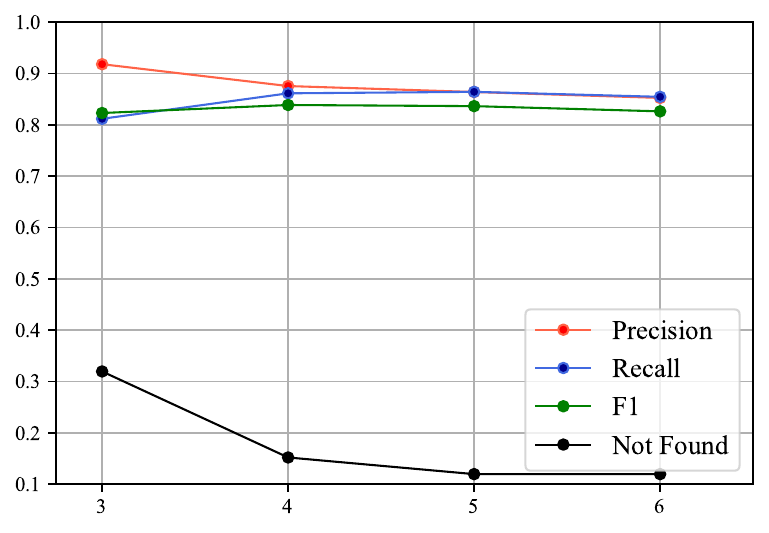}
\vspace{-5pt}
\caption{Network Depth Limit}\label{fig:depth}
\end{subfigure}
~
\begin{subfigure}[b]{0.5\textwidth}
\centering
\includegraphics[scale=0.55]{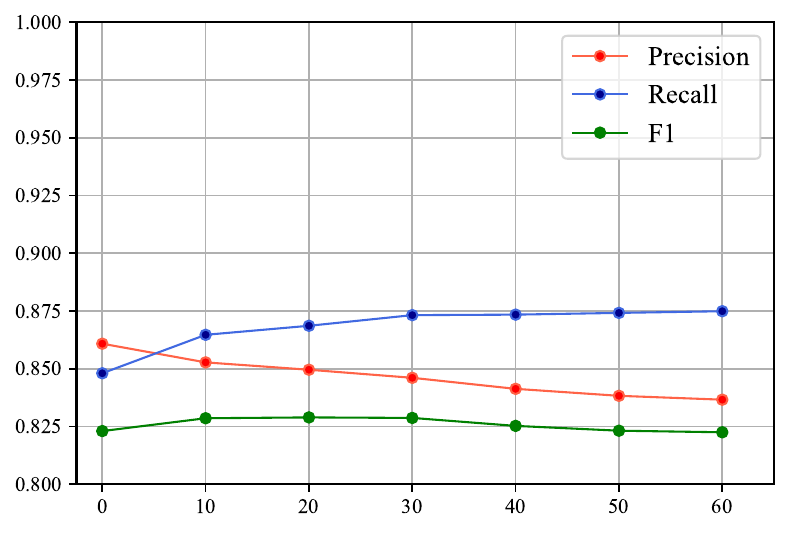}
\vspace{-5pt}
\caption{Commit Span}\label{fig:span}
\end{subfigure}
\vspace{-20pt}
\caption{Sensitivity Analysis Result for Network Depth Limit and Commit Span}\label{fig:sensitivity}
\end{figure*}

\tool has two configurable parameters, i.e., the network depth~limit in the first step of~\tool and the commit span~in~the~third step.~The default configuration is 5 and 30, which is used in the~evaluation~for \textbf{RQ6}, \textbf{RQ7}, \textbf{RQ8} and \textbf{RQ9}. To evaluate the sensitivity~of~\tool~to the two parameters, we reconfigured one parameter~and~fix~the~other, and reran \tool against our depth dataset.~Specifically, the network depth limit was configured from 3 to 6 by a step~of~1, and the commit span was configured \congyingEdit{from 0 to 60 by a step of 10}.

Fig.~\ref{fig:depth} and \ref{fig:span} show the impact of the two parameters on the~accuracy of \tool, where $x$-axis denotes the value of the parameter, and $y$-axis denotes the accuracy of \tool. Overall,~as~the network depth limit increases, more potential patches~are~included in our reference network. 
\congyingEdit{The number of CVEs that \tool~finds~no~patch and precision decrease, and recall and F1-score first increase~and~then decrease.}
Hence, we believe 5 is a good value~for the network depth limit. As the commit span increases, a wider scope of commits~are searched. 
\congyingEdit{Precision decreases, recall increases,~and F1-score first increases and then decreases.}
Notice that the number~of~CVEs \tool finds~no patch will not change and thus is not presented~in Fig.~\ref{fig:span}. Hence, we believe 30 is a good value for the commit span. These results indicate that the sensitivity of the accuracy of \tool to the two configurable parameters is acceptable.

\begin{table}[!t]
\centering
\small
\caption{Overlap between \tool and Ground Truth}\label{table:distribution}
\vspace{-10pt}
\begin{tabular}{|c|c|c|}
\noalign{\hrule height 1pt}
$P_{\tool}$ vs. $P_{GT}$ & Number of CVEs & Sum of Found Patches\\
\noalign{\hrule height 1pt}
$P_{\tool} = P_{GT}$ & 773 (59.7\%)    & 1,451	\\
$P_{\tool} \supset P_{GT}$ & 128 (9.9\%) & 708	\\
$P_{\tool} \subset P_{GT}$ &	139 (10.7\%) & 289 \\
$P_{\tool} \cap P_{GT} \not= \emptyset$ & 27 (2.1\%) & 134	\\
$P_{\tool} \cap P_{GT} = \emptyset$ & 73 (5.6\%) & 250 \\
$P_{\tool} = \emptyset$ & 155 (12.0\%)  & 0 \\\hline
Total & 1,295 & 2,832 \\
\noalign{\hrule height 1pt}
\end{tabular}
\end{table}

\section{Application Analysis}

\tool is configurable to meet different accuracy requirements~of~applications. For applications that need \textbf{high patch precision}, \tool can be configured to not construct the reference network in a layered way but simply use the direct references contained~in~the~four~advisory sources (i.e., skipping reference analysis in the first step). As shown by our ablation analysis in the submitted paper, this~is~actually the variant $v_1^5$, and achieves the highest precision of 0.918,~6.3\% higher than that of the original \tool. 

For applications that need \textbf{high patch recall}, \tool can~be~configured to not follow the patch selection step in \tool but select~all patches in our reference network. As shown by our ablation analysis in the submitted paper, this~is~actually the variant $v_2^1$,~and~achieves the highest recall of 0.940, 8.8\% higher~than~that~of~the~original~\tool. 

These results demonstrate that the two variants of \tool can meet the practical requirements of high precision and high recall.


\bibliographystyle{ACM-Reference-Format}
\bibliography{src/reference}
